\documentclass[11pt]{article}
\usepackage{psfig,amssymb,latexsym,eufrak,fancyheadings}

  \newlength\titlebox 
\newcommand{\Taver}[1]{  \left \langle #1 \right \rangle }
\newcommand{\Saver}[1]{  \left \langle \! \left \langle  #1
                        \right \rangle \! \right \rangle }

\newcommand{\sction}[1]{\section{#1 \hrulefill}}  

\title{ \textbf{Data clustering using a model granular magnet}}

\author{Marcelo Blatt, Shai Wiseman and Eytan Domany \\
        Department of Physics of Complex Systems, \\
        The Weizmann Institute of Science, 
        Rehovot 76100, Israel
}
\date{\today}
\begin{document}

\renewcommand{\baselinestretch}{1.50}
\tiny\normalsize

\maketitle

\begin{abstract}
  We present a new approach to clustering, based on the physical
  properties of an inhomogeneous ferromagnet.  No assumption is made
  regarding the underlying distribution of the data.  We assign a
  Potts spin to each data point and introduce an interaction between
  neighboring points, whose strength is a decreasing function of the
  distance between the neighbors.  This magnetic system exhibits three
  phases. At very low temperatures it is completely ordered; {\em
    i.e.} all spins are aligned. At very high temperatures the system
  does not exhibit any ordering and in an intermediate regime clusters
  of relatively strongly coupled spins become ordered, whereas
  different clusters remain uncorrelated.  This intermediate phase is
  identified by a jump in the order parameters. The spin-spin
  correlation function is used to partition the spins and the
  corresponding data points into clusters.  We demonstrate on three
  synthetic and three real data sets how the method works.  Detailed
  comparison to the performance of other techniques clearly indicates
  the relative success of our method.

\end{abstract}

\vspace*{\fill}

\newpage

\sction{Introduction}
In recent years there has been significant interest in adapting
numerical (Kirkpatrick, Gelatt and Vecchi 1983) as well as analytic (Fu
and Anderson 1986, M\'ezard and Pairsi 1986) techniques from statistical
physics to provide algorithms and estimates for good approximate
solutions to hard optimization problems (Yuille and Kosowsky 1994).  In
this work we formulate the problem of data clustering as that of
measuring equilibrium properties of an inhomogeneous Potts model. We are
able to give good clustering solutions by solving the physics of this
Potts model.

Cluster analysis is an important technique in exploratory data
analysis, where a priori knowledge of the distribution of the observed
data is not available (Duda and Hart 1973; Jain and Dubes 1988).  {\em
  Partitional} clustering methods, that divide the data according to
natural classes present in it, have been used in a large variety of
scientific disciplines and engineering applications that include
pattern recognition (Duda and Hart 1973), learning theory (Moody and
Darken 1989), astrophysics (Dekel and West 1985), medical image
(Suzuki {\em et.\ al.\ } 1995) and data processing (Phillips {\em et.\ 
  al.\ } 1995), machine translation of text (Cranias {\em et.\ al.\ }
1994), image compression (Karayiannis {\em et.\ al.\ } 1994),
satellite data analysis (Baraldi and Parmiggiani 1995), automatic
target recognition (Iokibe 1994), as well as speech recognition
(Kosaka and Sagayama 1994), and analysis (Foote and Silverman 1994).

The goal is to find a partition of a given data set into several compact
groups. Each group indicates the presence of a distinct category in the
measurements.  The problem of partitional clustering can be formally
stated as follows.  Determine the partition of $N$ given patterns $\left
  \{ v_i \right \}_{i=1}^N$ into groups, called {\em clusters}, such
that the patterns of a cluster are more similar to each other than to
patterns in different clusters. It is assumed that either $d_{ij}$, the
measure of dissimilarity between patterns $v_i$ and $v_j$ is provided,
or that each pattern $v_i$ is represented by a point $\vec x_i$ in a
$D$-dimensional metric space, in which case $d_{ij}=|\vec x_i-\vec
x_j|$.

The two main approaches to partitional clustering are called {\em
  parametric} and {\em non-parametric}. In parametric approaches some
knowledge of the clusters' structure is assumed and in most cases
patterns can be represented by points in a $D$-dimensional metric
space.  For instance, each cluster can be parameterized by a center
around which the points that belong to it are spread with a locally 
Gaussian distribution.  In many cases the assumptions are incorporated
in a {\em global criterion} whose minimization yields the ``optimal''
partition of the data.  The goal is to assign the data points so that
the criterion is minimized.  Classical approaches are variance
minimization, maximal likelihood and fitting Gaussian mixtures.  A
nice example of variance minimization is the method proposed by Rose,
Gurewitz and Fox (1990) based on principles of statistical physics,
which ensures an optimal solution under certain conditions.  This work
gave rise to other mean-field methods for clustering data (Buhmann and
K{\"u}hnel 1993, Wong 1993, Miller and Rose 1996). 
Classical examples of fitting Gaussian
mixtures are the Isodata algorithm (Ball and Hall 1967) or its
sequential relative, the K-means algorithm (MacQueen 1967) in
statistics, and soft competition in neural networks (Nowlan and Hinton,
1991).

In many cases of interest, however, there is no a priori knowledge
about the data structure. Then it is more natural to adopt
non-parametric approaches, which make less assumptions about the model
and therefore are suitable to handle a wider variety of clustering
problems.  Usually these methods employ a {\em local criterion},
against which some attribute of the local structure of the data is
tested, to construct the clusters.  Typical examples are hierarchical
techniques such as the agglomerative and divisive methods (see Jain
and Dubes 1988).
These algorithms suffer, however, from at least one of the following
limitations: (a) high sensitivity to initialization; (b) poor
performance when the data contains overlapping clusters; (c) inability
to handle variabilities in cluster shapes, cluster densities and
cluster sizes.  The most serious problem is the lack of cluster
validity criteria; in particular, none of these methods provides an
index that could be used to determine the most significant partitions
among those obtained in the entire hierarchy.  All these algorithms
tend to create clusters even when no natural clusters exist in the
data.

We introduced recently (Blatt, Wiseman and Domany 1996a 1996b 1996c) 
a new approach to clustering, based on the physical properties of a 
magnetic system.  
This method has a number of rather unique advantages: it
provides information about the different self organizing regimes of
the data; the number of ``macroscopic'' clusters is an {\it output} of
the algorithm; hierarchical organization of the data is reflected in
the manner the clusters merge or split when a control parameter (the
physical {\it temperature}) is varied.  Moreover, the results are
completely insensitive to the initial conditions, and the algorithm is
robust against the presence of noise. The algorithm is computationally
efficient; equilibration time of the spin system scales with $N$, the
number of data points, and is {\it independent of the embedding
  dimension D}.

In this paper we extend our work by providing demonstrating the efficiency 
and performance of the algorithm on various real-life problems. Detailed
comparison with other non-parametric techniques are also presented. 
The outline of the paper is as follows.
The magnetic model and thermodynamic definitions are introduced in
section \ref{sec:model}. A very efficient Monte Carlo method used for
calculating the thermodynamic quantities is presented in section
\ref{sec:SWmethod}. The clustering algorithm is described in section
\ref{sec:TheAlgorithm}. In section \ref{sec:Applications} we analyze synthetic
and real data to demonstrate the main features of the method,
and comparison of its performance with other techniques.

\sction{The Potts model}
\label{sec:model}
We now briefly describe the physics of Potts models and introduce
various concepts and thermodynamic functions that will be used in our
solution of the clustering problem.

Ferromagnetic Potts models have been extensively studied for many years
(see Wu 1982 for a review).  The basic spin variable $s$ can take one of
$q$ integer values: $s=1,2,...q$.  In a magnetic model the Potts spins
are located at points $v_i$ that reside on (or off) the sites of some
lattice.  Pairs of spins associated with points $i$ and $j$ are coupled
by an interaction of strength $J_{ij} >0$. Denote by $\cal S$ a
configuration of the system, ${\cal S} = \{ s_i \}^N_{i=1}$. The energy
of such a configuration is given by the Hamiltonian
\begin{equation}
     {\cal H}{\scriptstyle ({\cal S})} = 
     \sum_{ <i,j>} J_{ij} \; \left ( 1 - \delta_{ s_i, s_j} \right )
     \hspace{1.5cm} \mbox{$ s_i=1, \ldots q $} \;,
     \label{eq:Hamiltonian} 
\end{equation}
where the notation $<\! i,j \!>$ stands for neighboring sites $v_i$ and
$v_j$. The contribution of a pair $<\! i,j \! >$ to
${\cal H}$ is $0$ when $s_i=s_j$, {\em i.e.} when the two spins are aligned,
and is $J_{ij} > 0$ otherwise. If one chooses 
interactions that are a decreasing function of the distance
$d_{ij} \equiv  d \! \left (v_i, v_j \right)$, then the closer
two points are to each other, the more they ``like'' to be in
the same state.
 The Hamiltonian~(\ref{eq:Hamiltonian}) 
is very similar to other energy functions
used in neural systems, where each spin variable represents
a $q$-state neuron with an excitatory coupling to its neighbors. 
In fact, magnetic models have inspired many neural models (see for example
Hertz, Krogh and Palmer 1991).

In order to calculate the thermodynamic average of a physical quantity
$A$ at a fixed temperature $T$, one has to calculate the sum
\begin{equation} \label{eq:Taverage}
   \Taver{ A }=  \sum_{{\cal S}} \;
     A{\scriptstyle ({\cal S})} \; P{\scriptstyle ({\cal S})} \; ,
\end{equation}
where the Boltzmann factor
\begin{equation} \label{eq:Boltzmann}
     P{\scriptstyle ({\cal S})} \; = \;
     \frac{1}{Z} \;
     \exp \!\! \left (
         -\frac{ {\cal H}{\scriptscriptstyle ({\cal S})}}{T} 
     \right ) \; ,
\end{equation}
plays the role of the probability density which gives the statistical
weight of each spin configuration ${\cal S} = \left \{ s_i \right
\}_{i=1}^{N}$ in thermal equilibrium and $Z$ is a normalization constant,
$
  \label{eq:ZPotts}
  Z=\sum_{\cal S} \exp \! \left (
    -{\cal H}{\scriptscriptstyle ({\cal S})}/T
  \right )
$.

Some of the most important physical quantities $A$ for this magnetic
system are the order parameter or magnetization and the set of
$\delta_{s_i,s_j}$ functions, because their thermal average reflect
the ordering properties of the model.

The order parameter of the system is $\Taver{m}$, where the
magnetization, $m{\scriptstyle ({\cal S})}$, associated with a spin
configuration ${\cal S}$ is defined (Chen, Ferrenberg and Landau 1992)
as
\begin{equation}
  \label{eq:mag}
  m{\scriptstyle ({\cal S})}  = 
       \frac
           { q \, N_{\mbox{\scriptsize max}}{\scriptstyle ({\cal S})} - N}
           {(q-1)\,  N} \;
\end{equation}
with
\[
    N_{\mbox{\scriptsize max}}{\scriptstyle ({\cal S})}  =
    \mbox{max} 
       \left \{
             N_1{\scriptstyle ({\cal S})}, \;
             N_2{\scriptstyle ({\cal S})}, \ldots \;
             N_q{\scriptstyle ({\cal S})} 
       \right \} \; ,
\]
where $N_\mu{\scriptstyle ({\cal S})} $ is the number of spins with
the value $\mu$; $N_\mu{\scriptstyle ({\cal S})} = \sum_i
\delta_{s_i, \mu}$.

The thermal average of $\delta_{s_i,s_j}$ is called the spin--spin
correlation function,
\begin{equation}
  \label{eq:Gij}
   G_{ij} = \Taver{ \delta_{s_{i},s_{j}} } \; ,
\end{equation}
which is  the probability of the two spins $s_i$ and $s_j$  being aligned.

When the spins are on a lattice and all nearest neighbor couplings are
equal, $J_{ij}=J$, the Potts system is {\it homogeneous}. Such a model
exhibits two phases. At high temperatures the system is paramagnetic
or disordered; $\Taver{m}=0$, indicating that 
$N_{\mbox{\scriptsize max}}{\scriptstyle ({\cal S})} \approx \frac{N}{q}$
for all statistically significant configurations. In this phase the
correlation function $G_{ij}$ decays to $\frac{1}{q}$ when the distance
between points $v_i$ and $v_j$ is large; this is the probability to find
two completely independent Potts spins in the same state. At very high
temperatures even neighboring sites have $G_{ij} \approx \frac{1}{q}$.

As the temperature is lowered, the system undergoes a sharp transition
to an ordered, ferromagnetic phase; the magnetization jumps to
$\Taver{m} \neq 0$.  This means that in the physically relevant
configurations (at low temperatures) {\it one} Potts state ``dominates''
and $N_{\mbox{\scriptsize max}}{\scriptstyle ({\cal S})}$ exceeds
$\frac{N}{q}$ by a macroscopic number of sites. At very low temperatures
$\Taver{m} \approx 1$ and $G_{ij} \approx 1$ for all pairs $\{v_i,v_j\}$.

The variance of the magnetization is related to a relevant thermal
quantity, the susceptibility,
\begin{equation}
\label{eq:susceptibility}
  \chi =  \frac{N}{T} \left (\Taver{ m^2 } - \Taver{m}^2 \right ) \; ,
\end{equation}
which also reflects the thermodynamic phases of the system.
At low temperatures fluctuations of the magnetizations are negligible,
so the susceptibility $\chi$ is small in the ferromagnetic phase.

The connection between Potts spins and clusters of aligned spins
was established by Fortuin and Kasteleyn (1972). 
In the appendix 
we present such a relation and the probability distribution of such clusters.

We turn now to {\it strongly inhomogeneous} Potts models. This is the
situation when the spins form magnetic ``grains'', with very strong
couplings between neighbors that belong to the same grain, and very
weak interactions between all other pairs. At low temperatures such a
system is also ferromagnetic, but as the temperature is raised the
system may exhibit an intermediate, {\it super-paramagnetic} phase.
In this phase strongly coupled grains are aligned ({\em i.e.} are in
their respective ferromagnetic phases), while there is no relative
ordering of different grains.

At the transition temperature from the ferromagnetic to
super-paramagnetic phase a pronounced peak of $\chi$ is observed
(Blatt, Wiseman and Domany 1996a).  In the super-paramagnetic phase
fluctuations of the state taken by grains acting as a whole ({\em
  i.e.} as giant super-spins) produce large fluctuations in the
magnetization.  As the temperature is further raised, the
super-paramagnetic to paramagnetic transition is reached; each grain
disorders and $\chi$ abruptly diminishes by a factor that is roughly
the size of the largest cluster.  Thus the temperatures where a peak
of the susceptibility occurs and the temperatures at which $\chi$
decreases abruptly indicate the range of temperatures in which the
system is in its super-paramagnetic phase.

In principle one can have a sequence of several transitions in the 
super-paramagnetic phase: as the temperature is raised the system may break
first into two clusters, each of which breaks into more (macroscopic) 
sub-clusters and so on. Such a hierarchical structure of the magnetic
clusters reflects a hierarchical organization of the data into categories 
and sub-categories.

To gain some analytic insight into the behavior of inhomogeneous
Potts ferromagnets, we calculated the properties of such a ``granular'' 
system with
a macroscopic number of bonds for each spin. For such ``infinite range'' models
mean field is exact, and we have shown 
(Wiseman {\em et.\ al.\ } 1996, Blatt {\em et.\ al.\ } 1996b) that
in the paramagnetic phase
the spin state at each site is independent of any other spin; {\em i.e.}
$G_{ij} = \frac{1}{q}$.\\
At the paramagnetic/super-paramagnetic transition the correlation between 
spins belonging to the same group jumps abruptly to 
$
 \frac{q-1}{q} \left (\frac{q-2}{q-1} \right )^{\! 2} + \frac{1}{q}
 \simeq 1 - \frac{2}{q} + {\cal O }\!\left ( \frac{1}{q^2} \right )
$
while the correlation between spins belonging to different groups is unchanged.
The ferromagnetic phase is characterized by strong correlations between all 
spins of the system: 
$
   G_{ij} > \frac{q-1}{q} \left (\frac{q-2}{q-1} \right )^{\! 2} + \frac{1}{q}.
$ 

There is an important lesson to remember from this: in mean field we see that
in the super-paramagnetic phase two spins that belong to the same grain 
are strongly correlated, whereas for pairs that do not belong to the same grain
$G_{ij}$ is small. As it turns out, this double-peaked distribution of 
the correlations is {\it not an artifact} of mean field and will be used
in our solution of the problem of data clustering.

As we will show below, we use the data points of our clustering problem
as sites of an inhomogeneous Potts ferromagnet. Presence of clusters in
the data gives rise to magnetic grains of the kind described above
in the corresponding Potts model. Working in the super-paramagnetic phase of
the model we use the values of the pair correlation function of the Potts spins
to decide whether a pair of spins do or do not belong to the same grain and we
identify these grains as the clusters of our data. This is the
essence of our method; details are given below.

\section{Monte Carlo simulation of Potts models: the Swendsen-Wang method}
\label{sec:SWmethod}
The aim of equilibrium statistical mechanics is to evaluate sums such as
(\ref{eq:Taverage}) for models with $N >>1$  spins.
\footnote{Actually one is usually interested in the {\it thermodynamic limit}, 
e.g. when the number of spins $N \rightarrow \infty$}
This can be done analytically only for very limited cases. One resorts 
therefore to various approximations (such as mean field), or to computer 
simulations that aim at evaluating thermal averages numerically.

Direct evaluation of sums like (\ref{eq:Taverage}) is impractical, since  
the number of  configurations ${ \cal S}$ increases exponentially with
 the system size $N$.
Monte Carlo simulations methods (see Binder and Heermann 1988 for a
introduction) overcome this problem by generating a characteristic
subset of configurations which are used as a statistical sample.  They
are based on the notion of {\em importance sampling}, in which a set of
spin configurations
$\left \{ {\cal S}_1, {\cal S}_2, \dots {\cal S}_M \right \}$
is generated according to the Boltzmann probability distribution
(\ref{eq:Boltzmann}).  Then, expression (\ref{eq:Taverage}) is
reduced to a simple arithmetic average
\begin{equation} \label{eq:average}
    \Taver{A} \; \approx \; \frac{1}{M}
    \sum_i^M \;  A{\scriptstyle ({\cal S}_i)} 
\end{equation}
where the number of configurations in the sample, $M$, is much smaller
than $q^N$, the total number of configurations.  The set of $M$ states
necessary for the implementation of (\ref{eq:average}) are constructed
by means of a Markov process in the configuration space of the system.
There are many ways to generate such a Markov chain: in this work it
turned out to be essential to use the Swendsen--Wang (Wang and
Swendsen 1990, Swendsen, Wang and Ferrenberg 1992) Monte Carlo
algorithm (SW).  The main reason for this choice is that it is
perfectly suitable for working in the super-paramagnetic phase: it
overturns an aligned cluster in one Monte Carlo step, whereas
algorithms that use standard local moves will take forever to do this.

The first configuration can be chosen at random 
(or by setting all $s_i=1$).
Say we already generated $n$ configurations of the system, 
$\left \{ {\cal S}_i \right \}^n_{i=1}$,
and we start to generate configuration $n+1$. This is the way it is done. 

First ``visit'' all pairs of spins $<\! i,j \!>$ that interact, {\em i.e.}
have $J_{ij}>0$; 
the two spins are ``frozen'' together  
with probability 
\begin{equation} \label{eq:pf}
  p^f_{i,j} = 1 - 
        \exp \!\! 
          \left ( {-\frac{J_{ij}}{T}} \delta_{s_i,s_j}
         \right ) \;,
\label{eq:SWPd}
\end{equation} 
That is, if in our current configuration ${\cal S}_n$
the two spins are {\it in the same state}, $s_i=s_j$, then
sites $i$ and $j$ are frozen with probability
$p^f= \exp (-J_{ij}/T)$.

Having gone over all the interacting pairs, the next step 
of the algorithm is to identify the SW--clusters of spins. A SW--cluster
contains all spins which have a path of
frozen bonds connecting them.
Note that according to (\ref{eq:SWPd}) only spins of
the same value can be frozen in the same SW--cluster. After this step 
our $N$ sites are assigned to some number of distinct SW--clusters. If we
think of the $N$ sites as vertices of a graph whose edges are the 
interactions between neighbors $J_{ij}>0$, each SW--cluster is a subgraph
of vertices connected by frozen bonds.

The final step of the procedure is to generate the  
new spin configuration ${\cal S}_{n+1}$. This is done by
drawing, independently for each SW--cluster,
randomly a value $s=1,\ldots q$, which is assigned to all its spins.
This  defines one Monte Carlo step ${\cal S}_n \rightarrow {\cal S}_{n+1}$.
By iterating this procedure $M$ times while calculating at each
Monte Carlo step the physical quantity $A{\scriptstyle ({\cal S}_i)}$
the thermodynamic average  (\ref{eq:average}) is obtained.
The physical quantities that we are interested in are the magnetization
(\ref{eq:mag}) and its square value for the calculation of the 
susceptibility $\chi$, and
the spin--spin correlation function (\ref{eq:Gij}). Actually, in most
simulations a number of the early
configurations are discarded, to allow the system to ``forget''
its initial state. This is not necessary if 
the number of configurations $M$ is not too small (increasing $M$ improves
of course the statistical accuracy of the Monte Carlo measurement).  
Measuring autocorrelation times (Gould and Tobotchnik 1989)
provides a way 
of both deciding on the number of discarded configurations and for
checking that the number of configurations 
$M$ generated is sufficiently large.
A less rigorous way is simply plotting the
energy as a function of the number of SW steps and verifying 
that the energy reached a stable regime. 

At temperatures where large regions of correlated spins occur, local
methods  (such as Metropolis), which flip one spin at a time,
become very slow. The SW procedure overcomes this difficulty by
flipping  large clusters of aligned spins simultaneously.  Hence the
SW method exhibits much smaller autocorrelation times than local
methods.  The efficiency of the SW method, which is widely used in
numerous applications, has been tested in various
Potts (Billoire 1991) and Ising (Hennecke 1993) models.

\sction{Clustering of Data - detailed description of the algorithm}
\label{sec:TheAlgorithm}
So far we have
defined the Potts model, the various thermodynamic functions
that one measures for it and the (numerical) method used to measure
these quantities. We can now turn to the problem for which these concepts will 
be utilized, {\em i.e.} clustering of data. 

For the sake of concreteness assume that our data consists of $N$ patterns or
measurements $v_i$, 
specified by  $N$ corresponding vectors ${\vec x}_i$, embedded
in a $D$-dimensional metric space. Our method consists of three stages.  
The starting point is the
specification of the Hamiltonian (\ref{eq:Hamiltonian}) which governs
the system.  Next, by measuring the susceptibility $\chi$ and magnetization as 
function of temperature the different phases of the model are identified. 
Finally, the correlation
of neighboring pairs of spins, $G_{ij}$, is measured.  This correlation
function is then used to partition the spins and the
corresponding data points into clusters.

The outline of the three stages and the sub-tasks contained in each can
be summarized as follows:

\begin{enumerate}

   \item Construct the physical analog Potts-spin problem:
   \begin{enumerate} 
      \item Associate a Potts spin variable $s_i = 1, 2,\dots q$ to each
        point $v_i$.
       \item Identify the neighbors of each point $v_i$
         according to a selected criterion. 
       \item Calculate the interaction $J_{ij}$ between
         neighboring  points $v_i$ and  $v_j$.
      \end{enumerate}

      \item Locate the super-paramagnetic phase.
        \begin{enumerate} 
           \item Estimate the (thermal) average magnetization, 
             $\Taver{m}$, for different 
             temperatures. 
           \item Use the susceptibility $\chi$ to identify the
             super-paramagnetic phase.            
      \end{enumerate}          

      \item In the super-paramagnetic regime 
      \begin{enumerate} 
           \item Measure the spin--spin correlation, $G_{ij}$, 
             for all neighboring points $v_i$, $v_j$.
           \item Construct the data-clusters.
      \end{enumerate}   
\end{enumerate}

In the following subsections we provide detailed descriptions of the
manner in which each of the three stages are to be implemented.

\subsection{The physical analog Potts-spin problem.}
The goal is to specify the Hamiltonian 
of the form (\ref{eq:Hamiltonian}), that serves as the physical
analog of the data points to be clustered. One has to
assign a Potts spin to each data point, and
introduce short-range interactions between spins that reside on
neighboring points. Therefore we have to choose the 
value of $q$, the number of possible states a
Potts spin can take, define what is meant by ``neighbor'' points 
and provide the functional dependence of the interaction strength $J_{ij}$
on the distance between neighboring spins.

We discuss now the
possible choices for these attributes of the Hamiltonian
and their influence on the algorithm's performance.
The most important observation is that {\it none} 
of them needs fine tuning; the algorithm performs well provided a reasonable
choice is made, and the range of ``reasonable choices'' is very wide. 

\subsubsection{The Potts spin variables}
The number of Potts states, $q$, determines mainly the sharpness
of the transitions and the temperatures at which they occur.
The higher $q$, the sharper the transition.\footnote{
For a  two dimensional regular lattice one must have $q>4$ 
to ensure that the transition is of first order, in which case
the order parameter exhibits a discontinuity (Baxter 1973,  Wu 1982)}
On the other hand, in order to maintain a given statistical accuracy,
it is necessary to perform longer simulations
as the value the $q$ increases.
>From our simulations we conclude that the influence of $q$ on the
resulting classification is weak. We used $q=20$ in all the
examples presented in this work.
 
Note that the value of $q$ does not imply any assumption about the
number of clusters present in the data.

\subsubsection{Identifying Neighbors} The need for identification of
the neighbors of a point ${\vec x}_i$ could be eliminated by letting
all pairs $i,j$ of Potts spins interact with each other, via a short
range interaction $J_{ij} = f{\scriptstyle (d_{ij})}$ which decays
sufficiently fast (say, exponentially or faster) with the distance
between the two data points.  The phases and clustering properties of
the model will not be affected strongly by the choice of $f$. Such a
model has ${\cal O}(N^2)$ interactions, which makes its simulation
rather expensive for large $N$. For the sake of computational
convenience we decided to keep only the interactions of a spin with a
limited number of neighbors, and setting all other $J_{ij}$ to zero.
Since the data do not form a regular lattice, one has to supply some
reasonable definition for ``neighbors''.  As it turns out, our results
are quite insensitive to the particular definition used.

Ahuja (1982) argues for intuitively appealing
characteristics of Delaunay triangulation over other graphs structures
in data clustering. We use this definition when the patterns are
embedded in a low dimensional ($D \le 3$) space.

For higher dimensions we use the {\em mutual neighborhood value}; we
say that $v_i$ and $v_j$ have a mutual neighborhood value $K$, if and
only if $v_i$ is one of the K-nearest neighbors of $v_j$ and $v_j$ is
one of the K-nearest neighbors of $v_i$.
We chose $K$ such that the interactions connect all data points to one
connected graph.
Clearly $K$ grows with the dimensionality.
We found convenient, in cases of very high dimensionality ($D > 100$),
to fix $K = 10$ and to superimpose to the edges obtained with this criteria
the edges corresponding to the minimal spanning tree associated with the
data. We use this variant only in the examples presented in 
sections \ref{sec:one.example} and
\ref{sec:two.gauss.example}.

\subsubsection{Local interaction}
\label{sec:interaction}
In order to have a model with the physical properties of a strongly
inhomogeneous granular magnet, we want strong interaction between
spins that correspond to data from a high-density region, and weak 
interactions between neighbors that are in low-density regions.
To this end and 
in common with
other ``local methods'', we assume that there is a `local length
scale' $\sim a$, which is defined by the high density regions and is
smaller than the typical distance between points in the low density
regions.  This $a$ is the characteristic scale over which our short-range 
interactions decay. We tested various choices but report here only
results that were obtained using 
 
\begin{equation} \label{eq:J}
   J_{ij} =  \left \{
        \begin{array}{ll}
            \frac{1}{\widehat{K}} \; \exp \! \!  \left (
                - \frac{d_{ij}^2}{2 a^2} \right ) & 
                       \mbox{if $v_i$ and $v_j$ are neighbors} \\
                   0 & \mbox{otherwise}
        \end{array}
      \right .
\end{equation}

We chose the ``local length scale'', $a$, to be the average of all
distances $d_{ij}$ between neighboring pairs $v_i$ and $v_j$.
$\widehat K$ is the average number of neighbors; it is twice the
number of non vanishing  interactions divided by the
number of points $N$.  This careful normalization of the interaction
strength enables us to estimate the temperature corresponding to
the highest super-paramagnetic transition 
(see \ref{sec:super-paramagnetic}).

It should be noted that everything done so far can be easily implemented
in the case when instead of providing the ${\vec x}_i$ for all the data
we have an $N \times N$ matrix of dissimilarities $d_{ij}$. This was 
tested 
in experiments for clustering of images
where only a measure of the dissimilarity between them was available 
(Gdalyahu and Weinshall 1996). Application of other clustering
methods would have necessitated  
embedding  this
data in a metric space; the need for this was eliminated by using SPC. 
The results obtained by applying the method on the matrix of 
dissimilarities\footnote{Interestingly, the triangle inequality was violated
in about 5\% of the cases.}
of these images were excellent; all points were classified with no error.

\subsection{Locating the super-paramagnetic regions}
\label{sec:super-paramagnetic}
The various temperature intervals in which the system
self-organizes into different partitions to clusters
are identified by measuring the susceptibility $\chi$
as a function of temperature. 
We start by summarizing the Monte Carlo procedure and conclude by
providing an estimate of the 
highest transition temperature to the super-paramagnetic regime. Starting from 
this estimate, one can take increasingly refined temperature scans and 
calculate the function $\chi (T)$ by Monte Carlo simulation.

We used the Swendsen--Wang method described in section \ref{sec:SWmethod};
here we give a step by step summary of the procedure.

\begin{enumerate}
   \item Choose the number of iterations $M$ to be performed.
   \item Generate the initial 
     configuration by assigning a random value to 
     each spin.
   \item  Assign frozen bond between nearest neighbors points $v_i$ and  $v_j$ 
     with probability $p^f_{i,j}$ (eq. (\ref{eq:pf})).
   \item Find the connected subgraphs, the SW--clusters.  
   \item Assign new random values to the spins (spins that belong to the same 
     SW--cluster are assigned the same value). This is the new configuration
      of the system.
   \item Calculate the value assumed by the physical quantities of interest
     in the new spin configuration.
   \item Go to step 3, unless the maximal number of iterations $M$, was reached.
   \item Calculate the averages (\ref{eq:average}).
\end{enumerate}

The super-paramagnetic phase can contain many different
sub-phases with different ordering properties. A typical example
can be generated by data with a hierarchical structure, giving rise to
different acceptable partitions of the data. 
We measure the susceptibility $\chi$ at different temperatures
in order to locate these different regimes.
The aim is to identify the temperatures at which the system
changes its structure.

As was discussed in section \ref{sec:model}, the
super-paramagnetic phase is characterized by a non-vanishing
susceptibility. Moreover, there are two basic features of $\chi$ in
which we are
interested. The first is a peak in the susceptibility, which signals
a ferromagnetic
to super-paramagnetic transition, at which a large cluster breaks  
into a few smaller (but still macroscopic) clusters.
The second feature is an abrupt decrease of the susceptibility,
corresponding to a super-paramagnetic to paramagnetic transition, in which
one or more large clusters have melted 
({\em i.e.} broke up into many {\it small} clusters). 

The location of the super-paramagnetic to paramagnetic transition
which occurs at the highest
temperature can be  roughly estimated by the following
considerations.  First we approximate the clusters by an ordered
lattice of coordination number  $\widehat K$ and a constant
interaction 
\[
   J  \approx \Saver{ J_{ij} } =
   \Saver{ \frac{1}{\widehat K} 
      \exp \!\! \left ( -\frac{ d_{ij}^2 }{ 2 a^2 } \right ) }
   \; \approx \;
   \frac{1}{\widehat K} \exp \!\! \left (
            -\frac{ \Saver{d_{ij}^2} }{ 2 a^2 } 
            \right ) \;
\]
where $\Saver{\cdots}$ denotes the average over all neighbors.
Secondly, from the
Potts model on a square lattice (Wu 1982), we get that this transition should
occur roughly at
\begin{equation} \label{eq:T_est}
 T \approx \frac{1}{4 \log(1 + \sqrt{q})}
              \exp \!\! \left (
                   -\frac{ \Saver{d_{ij}^2} }{ 2 a^2 } 
              \right ) \; .
\end{equation}
An estimate based on the mean field model yields a very similar
value.

\subsection{Identifying the data clusters}
\label{sec:identify}
Once the super-paramagnetic phase and its
different sub-phases have been identified, we select {\it one
temperature} in each region of interest. The rational is that each sub-phase
characterizes a particular type of partition of the data, 
with new clusters merging or breaking. On the other hand, 
as the temperature is varied 
{\it within} a phase, one expects only shrinking or expansion of
the existing clusters,  
changing only the classification 
of the 
points on the
boundaries of the clusters.

\subsubsection{The spin--spin correlation}
\label{sec:spinspin}
We use the spin--spin correlation function $G_{ij}$, between neighboring
sites $v_i$ and $v_j$, to build the 
data clusters.
In principle we have to calculate the thermal average (\ref{eq:average})
of $\delta_{s_i,s_j}$ in order to obtain  $G_{ij}$. 
However, the Swendsen--Wang method provides an improved
estimator (Niedermayer 1990) of the
spin--spin correlation function. One 
calculates the two--point connectedness $C_{ij}$, the probability that
sites $v_i$ and $v_j$ belong to the same SW--cluster, which
is estimated by the average
(\ref{eq:average}) of the following indicator function
\[
  c_{ij} = \; \left \{
     \begin{array}{ll}
        1  & \mbox{if $v_i$ and $v_j$ belong to the same SW--cluster}\\
        0  & \mbox{otherwise} 
      \end{array}
   \right . \; ,
\]
$C_{ij} = \Taver{c_{ij}}$ is the probability of finding 
sites $v_i$ and $v_j$ in the same SW--cluster.
Then the relation (Fortuin and Kasteleyn 1972)
\begin{equation}
    G_{ij} \; = \;
    \frac{ (q-1) \; C_{ij} +1 }{q} \; ;
  \label{eq:correlation}
\end{equation}
is used to obtain the correlation function $G_{ij}$.

\subsubsection{The  data clusters}
Clusters are identified in three steps. 
\begin{enumerate}
\item
Build the clusters'
``core'' using a thresholding procedure; if $G_{ij} > 0.5$, a link
is set between the neighbor data points $v_i$ and $v_j$.
The resulting connected graph
depends weakly on the value (0.5) used in this thresholding,
 as long as it is bigger than
$\frac{1}{q}$ and less than $1-\frac{2}{q}$. The reason is, 
as was pointed out in
section \ref{sec:model}, that the distribution of the
correlations between two neighboring
spins peaks strongly at these two values
and is very small between them (fig \ref{fig:aros_hist}$(b)$).

\item
Capture points lying on the periphery of the clusters
by linking each point $v_i$ to its neighbor $v_j$ of maximal
correlation $G_{ij}$.  
It may happen, of course, that points $v_i$ and $v_j$
were already linked in the previous step.  

\item
Data clusters are identified
as the linked components of the  graphs obtained in steps 1,2.
\end{enumerate}

Although it would be completely equivalent to use in steps 1,2 the
two--point connectedness, $C_{ij}$, instead of the 
spin--spin correlation, $G_{ij}$, we considered the latter
to stress the relation of our
method with the physical analogy we are using.

\newpage
\sction{Applications}
\label{sec:Applications}
The approach presented in this paper has been successfully tested on a
variety of data sets. The six examples we discuss here were chosen
with the intention of demonstrating the main features and utility of
our algorithm to which we refer from now as the 
{\em Super-Paramagnetic} clustering (SPC) method.  
We use both artificial and real data.  Comparisons with
the performance of other classical (non-parametric) methods are also 
presented. 
We refer to different clustering methods by the nomenclature 
used in the books of Jain and Dubes (1988) and Fukunaga (1990).

The non-parametric algorithms we have chosen belong to four families:
$a)$  hierarchical methods: {\em single--link} and
 {\em complete--link};
$b)$ graph theory based methods: {\em Zhan's minimal spanning tree} and 
Fukunaga's {\em directed graph method};
$c)$ nearest-neighbor clustering type, based on different proximity
measures: the {\em mutual neighborhood clustering} algorithm and 
{\em k-shared neighbors}; and 
$d)$ density estimation: Fukunaga's {\em valley seeking method}.
These algorithms are of the same kind as the super-paramagnetic
method in the sense that
only weak assumptions  are required about the underlying data structure.

The results from all these methods depend on various parameters in an
uncontrolled way;  we always  used the {\em best} 
result that was obtained. 

A unifying view of some of these methods in the framework of the 
present work is presented in the Appendix.

\subsection{A pedagogical 2-dimensional example}
\label{sec:2d.example}
The main purpose of this simple example is to illustrate the features of 
the method
discussed in the previous sections, in particular the behavior of the 
susceptibility and its use for the identification of the two 
kinds of phase transitions. The influence of the number of Potts states, $q$
and the partition of the data as a function of the temperature are also 
discussed.\\

\begin{figure}[htbp]
  \begin{center}
    \leavevmode
    \centerline{
      \psfig{figure=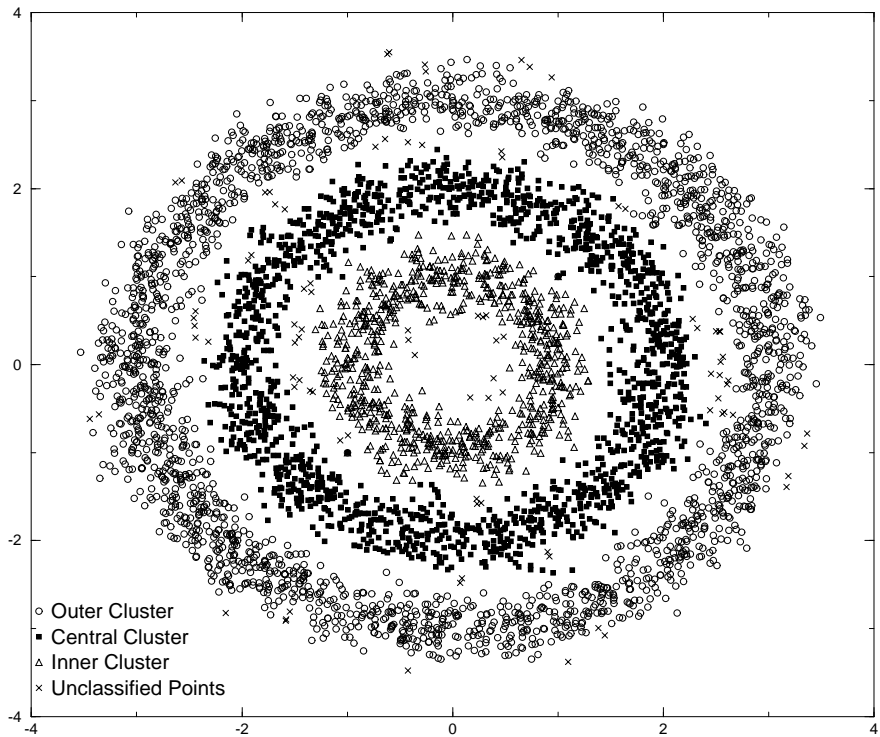,width=13.0cm,clip=}
      }
    \caption{{\em Data distribution}: the angular
      coordinate is uniformly distributed, {\em i.e} $\mbox{U}[0,2 \pi]$,
      while the radial one is normal $\mbox{N}[R,0.25]$ distributed around
      three different radius $R$. The outer cluster ($R=3.0$) consists
      of 2400 points, the central one ($R=2.0$) of 1600 and the inner
      one ($R=1.0$) of 800.  The classified data set: points
      classified at $T=0.05$ as belonging to the three largest
      clusters are marked by circles (outer cluster, 2358 points),
      squares (central cluster, 1573 points) and triangles (inner
      cluster, 779 points). The x's denotes the 90 remaining points
      which are distributed in 43 clusters, the biggest of size 4.}
    \label{fig:aros}
  \end{center}
\end{figure}

The toy problem of figure \ref{fig:aros} consist of 4800
points in $D=2$ dimensions whose angular distribution is uniform and
whose radial distribution is normal with variance $0.25$;
\begin{eqnarray*}
 \theta & \sim & \mbox{U}[0, 2\pi] \\
 r      & \sim & \mbox{N}[R, 0.25] \; ,
\end{eqnarray*}
we generated half the points with $R=3$,
one third with $R=2$ and one sixth with $R=1$.

Since there is a small overlap between the clusters, we consider the Bayes
solution as the optimal result; {\em i.e.} points whose distance to the 
origin is bigger than 2.5 are considered a cluster, points whose radial
coordinate lies between 1.5 and 2.5 are assigned to a second cluster
and the remaining points define the third cluster. These optimal clusters
consist of 2393, 1602 and 805 points respectively.

\begin{figure}[htbp]
  \begin{center}
    \leavevmode
    \psfig{figure=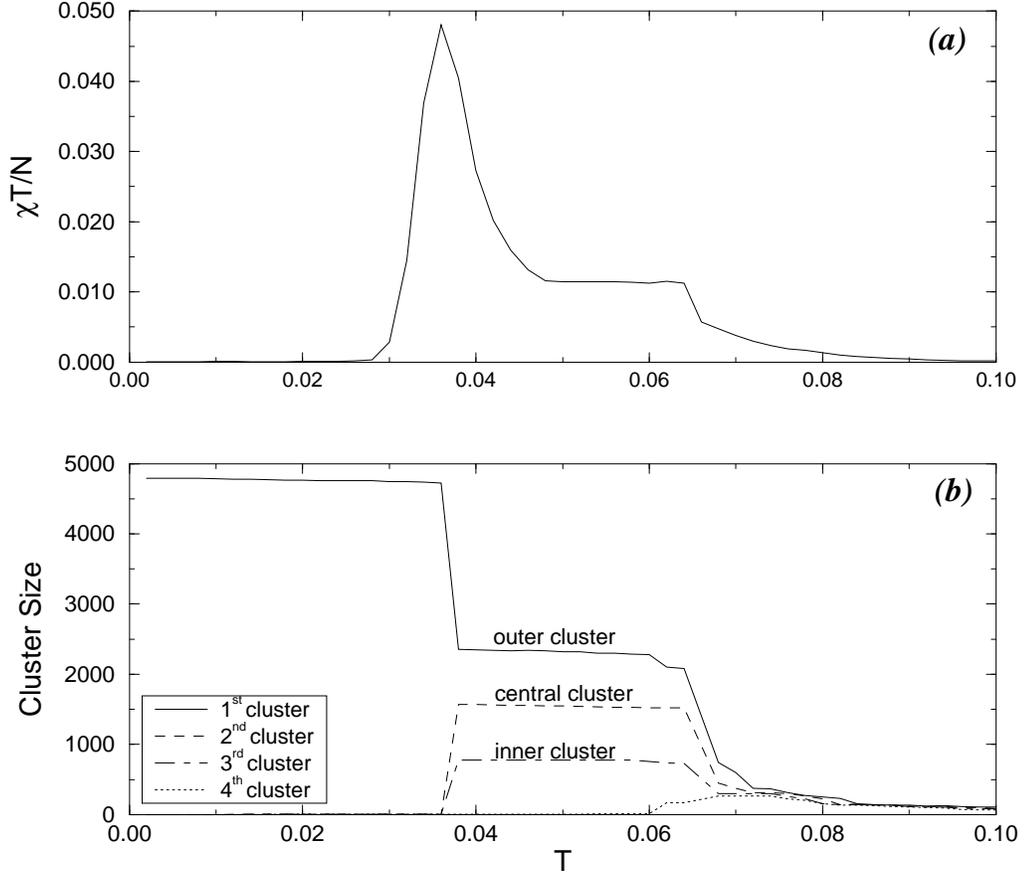,width=15.0cm}    
    \caption{$(a)$ The susceptibility density $\frac{\chi T}{N}$ of
      the data set of figure \protect{\ref{fig:aros}} as a
      function of the temperature.  $(b)$ Size of the four biggest
      clusters obtained at each temperature.}
    \label{fig:aros_TEMP}
  \end{center}
\end{figure}

By applying our procedure, and choosing the neighbors according to the
mutual neighborhood criterion with $K=10$ we obtain the
susceptibility as a function of the temperature as presented in figure
\ref{fig:aros_TEMP}$(a)$.  The estimated 
temperature (\ref{eq:T_est}) 
corresponding to the super-paramagnetic to paramagnetic
transition is 0.075, which is in a good agreement with the one inferred
from figure \ref{fig:aros_TEMP}$(a)$.

\vskip \baselineskip

Figure \ref{fig:aros} presents the clusters obtained at $T=0.05$.
The sizes of the three largest clusters are 2358, 1573 and 779
including 98\% of the data;
the classification of all these points coincides with that of the
optimal Bayes classifier. The remaining 90 points
are distributed among 43 clusters of size smaller than 4.
As can be noted in figure \ref{fig:aros}, the small clusters
(less than 4 points) are located at the boundaries between the main clusters. 

\begin{figure}[htbp]
  \begin{center}
    \leavevmode
    \hspace{\fill}
    \psfig{figure=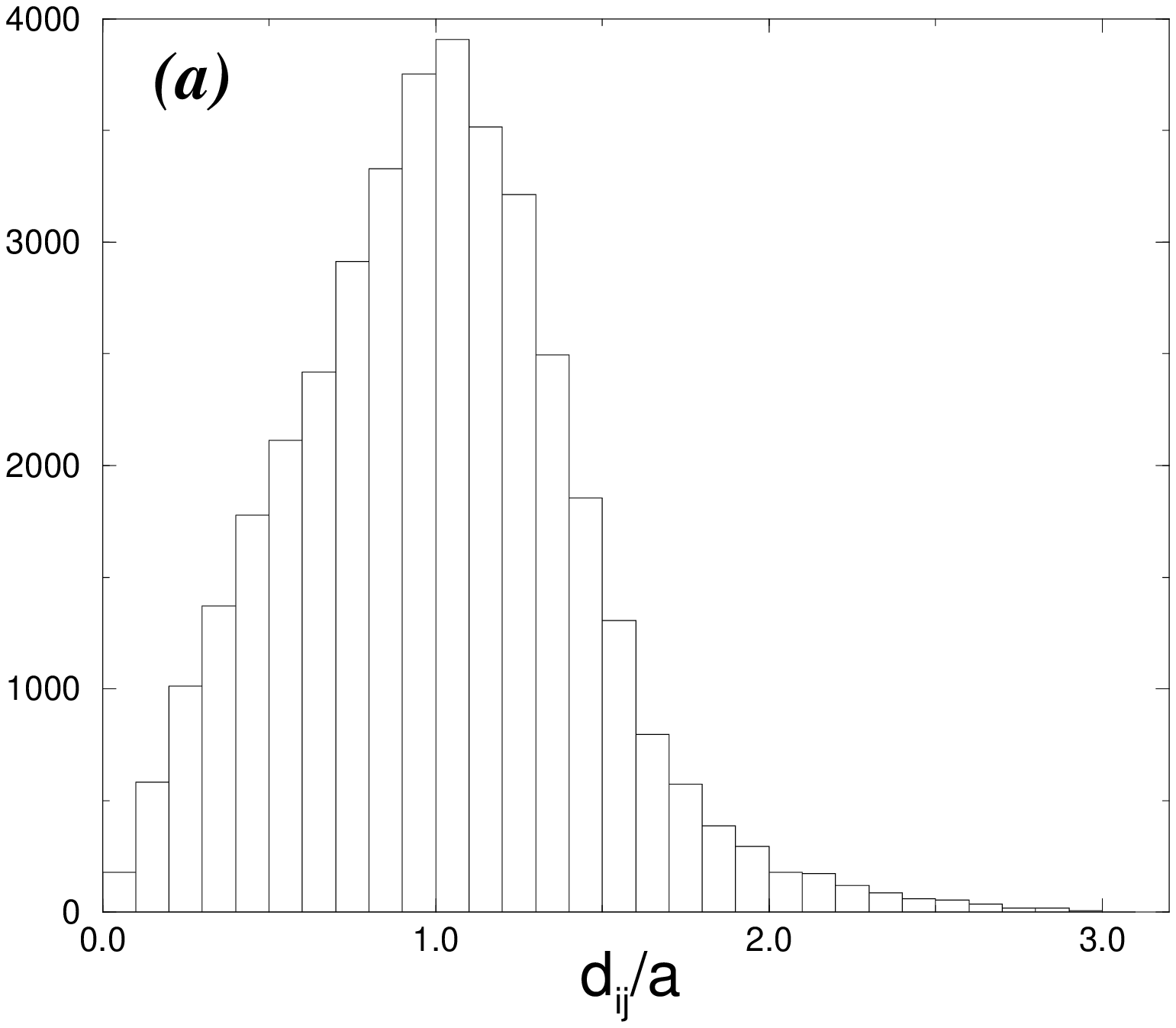,width=7.0cm} 
    \hspace{\fill}       
    \psfig{figure=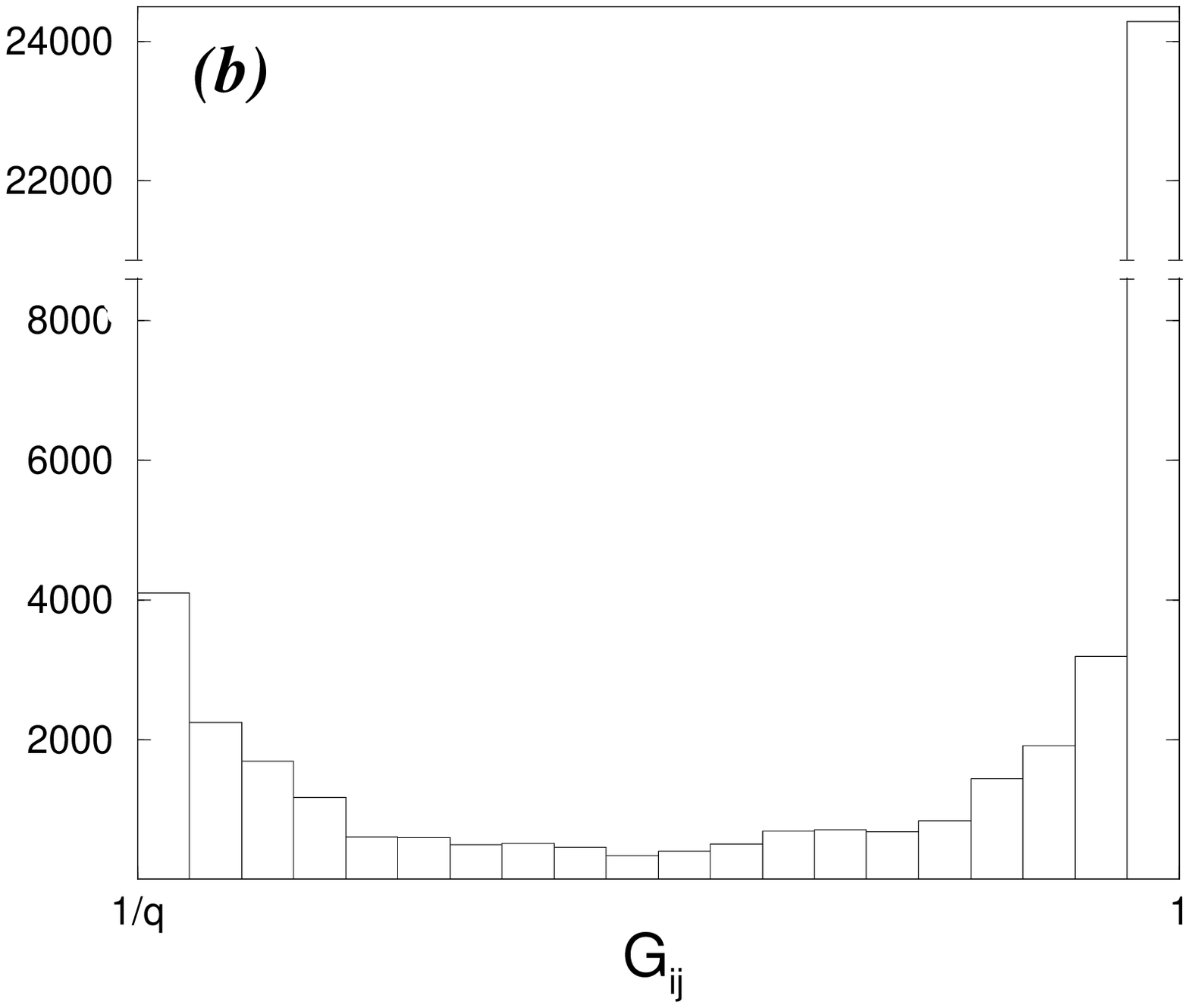,width=7.0cm}
    \hspace{\fill}
    \caption{Frequency distribution of $(a)$ distances between neighboring
      points of fig. \protect{\ref{fig:aros}} (scaled by the average 
      distance $a$), and $(b)$ spin--spin correlation of neighboring points.
      }
    \label{fig:aros_hist}
  \end{center}
\end{figure}

One of the most salient features of the SPC
method is that the spin--spin
correlation function, $G_{ij}$ reflects the existence of two categories
of neighboring points: neighboring points that belong  to the 
same cluster and those that do not. This  can be observed from 
figure  \ref{fig:aros_hist}$(b)$,
the two-peaked frequency distribution of the 
correlation function $G_{ij}$ between neighboring points of 
figure \ref{fig:aros}.
In contrast, the frequency distribution \ref{fig:aros_hist}$(a)$
of the  normalized distances $\frac{d_{ij}}{a}$ between neighboring 
points of figure \ref{fig:aros},  contains no hint 
of the existence of a natural
cutoff distance, that separates neighboring points into two categories.

It is instructive to observe the behavior of the size of the clusters
as a function of the temperature, presented in figure
\ref{fig:aros_TEMP}$(b)$. At low temperatures, as expected, all data
points form only one cluster.  At the ferromagnetic to
super-paramagnetic transition temperature, indicated by a peak in the
susceptibility, this cluster splits into three.  These essentially
remain stable in their composition until the super-paramagnetic to
paramagnetic transition temperature is reached, expressed in a sudden
decrease of the susceptibility $\chi$, where the clusters melt.

Turning now to the effect of the parameters on the procedure, we found
(Wiseman {\em et.\ al.\ } 1996) that the number of Potts states $q$
affects the sharpness of the transition but the obtained
classification is almost the same. For instance, choosing $q=5$ we found
that the three largest clusters contained 2349, 1569 and 774 data
points, while taking $q=200$ we yielded 2354, 1578 and 782.

Of all the algorithms listed at the beginning of this section, only 
the single--link
and minimal spanning methods were able to give (at the optimal values
 of their 
clustering parameter) a partition that reflects the
underlying distribution of the data. The best results are
summarized in table 
\ref{tab: result_aros}, together with those of the SPC method.
 Clearly, the standard parametric methods (such k-means or Ward's method)
would not be 
able to give a reasonable answer because they assume that different
clusters are parameterized by different centers and a spread around them.

\begin{table}[htbp]
  \begin{center}
    \leavevmode
    \begin{tabular}[htbp]{| l | r | r | r | r |}
      \hline
      \hline
      Method    &  ~~outer &  ~~~~central & ~~~~~~inner & ~~~~unclassified \\
                &  cluster &  ~~~~cluster & ~~~~cluster & ~~~~~~~~~~points \\
      \hline
      Bayes                        &    2393   &  1602   &  805   &    --  \\
      Super--Paramagnetic (q=200)  &    2354   &  1578   &  782   &    86  \\
      Super--Paramagnetic (q=~20)  &    2358   &  1573   &  779   &    90  \\
      Super--Paramagnetic (q=~~5)  &    2349   &  1569   &  774   &   108  \\
      Single--Link              &    2255   &  1513   &  758   &   274  \\
      Minimal Spanning Tree        &    2262   &  1487   &  756   &   295  \\
     \hline
     \hline     
    \end{tabular}
    \caption{Clusters obtained with the methods that succeeded in recovering
      the structure of the data. Points belonging to cluster of sizes
      less than 50 points are considered as ``unclassified points''.
      The Bayes method is used as benchmark because it is the one that
      minimizes the expected number of mistakes, provided that the
      distribution that generated the set of points is known }
    \label{tab: result_aros}
  \end{center}
\end{table}

\begin{figure}[htb]
  \begin{center}
    \leavevmode
    \psfig{figure=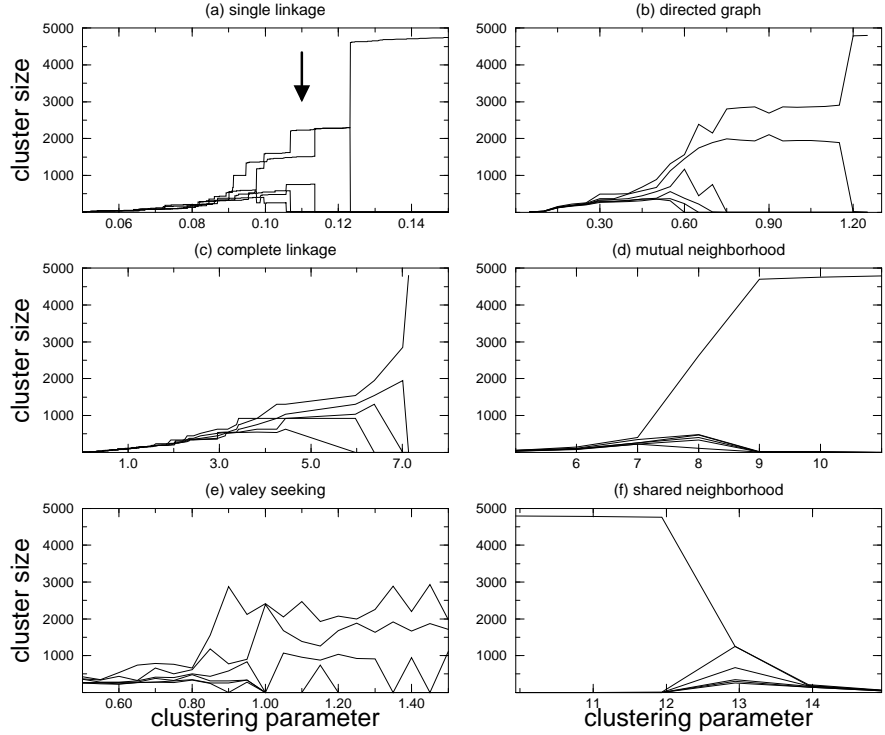,width=13.0cm}    
    \caption{Size of the three biggest clusters as a function of the
      clustering parameter obtained with $(a)$ single--link, $(b)$
      directed graph, $(c)$ complete--link, $(d)$ mutual neighborhood,
      $(e)$ valley seeking and $(f)$ shared neighborhood algorithm.
      The arrow in $(a)$ indicate the region corresponding to the
      optimal partition for the single--link method. The other
      algorithms were unable to recover the data structure.}
    \label{fig:aros_OTHER}
  \end{center}
\end{figure}

In figure \ref{fig:aros_OTHER} we present, for the methods that depend
only on a single parameter, the sizes of the four biggest
clusters that were obtained as a function of the clustering parameter.
The best solution  obtained with the single--link method
(for a narrow range of the parameter) corresponds also to three
big clusters of  2255, 1513 and 758 points respectively, while
the remaining clusters are of size smaller than 14. 
For larger threshold distance the 
second and third clusters are linked. This classification is slightly 
worse than the one obtained by the super--paramagnetic method.
 
When comparing SPC with single--link one should note that if the
``correct'' answer is not known, one has to rely on measurements such
as the stability of the largest clusters (existence of a plateau) to
indicate the quality of the partition.  As can be observed from figure
\ref{fig:aros_OTHER}$(a)$ there is no clear indication that signals
which plateau corresponds to the
optimal partition among the whole hierarchy yielded by single link. 
The best result obtained with the minimal spanning tree method is very
similar to the one obtained with the single--link, but this solution
corresponds to a very small fraction of its parameter space.
In comparison, SPC allows clear identification of the relevant 
super--paramagnetic phase; the entire temperature range of this regime
yields excellent clustering results.

\subsection{Only one cluster}
\label{sec:one.example}
Most existing algorithms impose a partition on the data even when there
are no natural classes present in it.  The aim of this example
is to show how the SPC algorithm signals this situation. Two different
100--dimensional data sets of 1000 samples are used. The first data set
is taken from a Gaussian distribution centered at the origin, with
covariance matrix equal to the identity. The second data set consists
of  points generated randomly from a uniform distribution
in a hypercube of side 2.

\begin{figure}[htbp]
  \begin{center}
    \leavevmode
    \psfig{figure=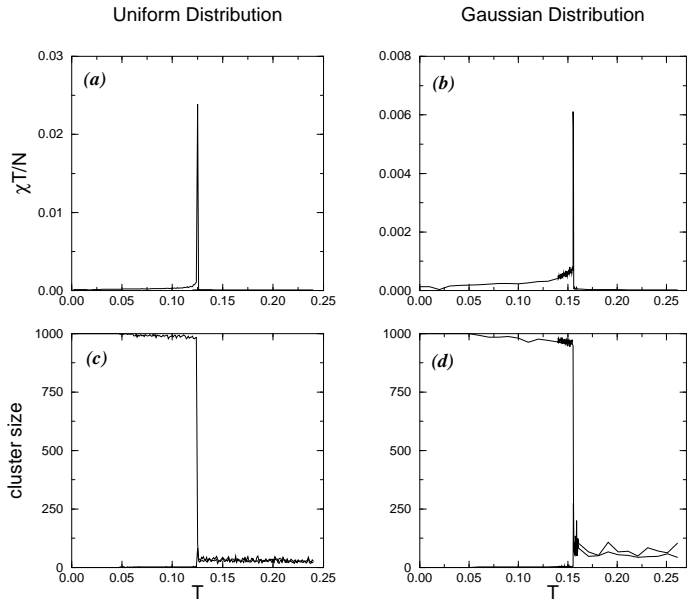,width=10.0cm}
    \caption{Susceptibility density $\frac{\chi T}{N}$ as a function of the
      temperature T for data points $(a)$ uniformly distributed in a
      hypercube of side 2 and $(b)$ multi-normally distributed with covariance
      matrix equal to the identity in a 100-dimensional space. The
      size of the two biggest clusters obtained at each temperature 
      are presented in $(c)$ and $(d)$
      respectively.}
    \label{fig:one_TEMP}
  \end{center}
\end{figure}

The susceptibility curve, which was obtained by using the SPC method
with these data sets is shown in figures \ref{fig:one_TEMP}$(a)$ and
\ref{fig:one_TEMP}$(b)$.
The narrow peak and the absence of a plateau indicate that there is
only a single phase transition (ferromagnetic to paramagnetic), with no
super-paramagnetic phase.  This single phase transition is
also evident from figures \ref{fig:one_TEMP}$(c)$ and
\ref{fig:one_TEMP}$(d)$ where only one cluster of almost 1000 points
appears below the transition.
This single ``macroscopic'' cluster ``melts'' at the transition,
to many ``microscopic'' clusters of 1--3 points in each. \\

Clearly, all existing methods {\em are} able to give the correct
answer since it is always possible to set the parameters such that
this trivial solution is obtained.  Again, however, there is no clear
indicator for the correct value of the control parameters of the
different methods.

\subsection{Performance: scaling with data dimension and influence of 
  irrelevant features}
\label{sec:two.gauss.example}
The aim of this example is to show the robustness of the SPC method
and to give an idea of the influence of the dimension of the data
on its performance.
To this end we generated $N$ $D$--dimensional points 
whose density distribution is 
a mixture of two isotropic Gaussians, {\em i.e.}

\begin{equation}
\label{eq:two_gauss}
   {\cal P}_{( \vec x )} = 
   \frac{\left ( \sqrt{2 \pi} \sigma \right )^{-D}}{2}
   \left [
     \exp \! \left ( - \frac{ \| \vec x - \vec y_1 \|^2}{2 \sigma^2} \right )+
     \exp \! \left ( - \frac{ \| \vec x - \vec y_2 \|^2}{2 \sigma^2} \right )
   \right ]
\end{equation}

where $\vec y_1$ and $\vec y_2$ are the centers of the Gaussians and
$\sigma$ determines its width.  Since the two characteristics lengths
involved are $\| \vec y_1 - \vec y_2 \|$ and $\sigma$, the relevant
parameter of this example is the normalized distance $L \; = \;
\frac{\| \vec y_1 - \vec y_2 \|}{\sigma}$.

The manner in which these data points were generated satisfies
precisely the hypothesis about data distribution that is assumed by
the K-means algorithm. Therefore it is clear that this
algorithm (with K=2) will achieve the Bayes optimal
result; the same will hold for other parametric methods, such as maximal 
likelihood (once a two-Gaussian distribution for the data is assumed).
Even though such algorithms have, for this kind of data, 
an obvious advantage over SPC, it is interesting to get a feeling about 
the loss in the quality of the results, caused by using our
method, which relies on less assumptions. To this end we considered the 
case of  $4000$ points generated in a $200$--dimensional space from the
distribution (\ref{eq:two_gauss}), setting the parameter $L = 13.0 \; \sigma$.
The two biggest clusters we obtained were of 
sizes $1853$ and $1816$; the smaller ones contained less than 4 points each.
About 8.0 \% of the points were left unclassified, but {\it all} 
those points that 
the method did assign to one of the two large clusters, were classified in 
agreement with a Bayes classifier. For comparison
we applied the single linkage algorithm to the same data;
at the {\em best} classification point
74\% of the points were unclassified. 

Next we studied
the minimal distance, $L_c$, at which the method is able to
recognize that two clusters are present in the data and, to
find the dependence of $L_c$ on
the dimension $D$ and number of samples $N$.  
Note that the lower
bound for the minimal discriminant distance for any non-parametric
algorithm is $2$ (for any dimension $D$). Below this distance the
distribution is no longer bimodal,
but rather the maximal density of points is located at the at the midpoint
between the Gaussian centers.
Sets of $N = 1000,
2000,4000$ and $8000$ samples and space dimensions
$D=2,10,100,100$ and $1000$ were tested.  We set the number of
neighbors $K=10$ and superimposed the minimal spanning tree to ensure
that at $T=0$ all points belong to the same cluster.  To our surprise
we observed that in the range $1000 \le N \le 8000$ the critical
distance seems to depend only weakly on the number of samples, $N$.
The second remarkable result is that the critical discriminant
distance $L_c$ grows very slowly with the dimensionality of the data
points, $D$.  Apparently the minimal discriminant distance $L_c$ 
increases like the logarithm of the
number of dimensions $D$;
\begin{equation}
\label{eq:Lc}
  L_c \approx \alpha + \beta \log D 
\end{equation}
where $\alpha$ and $\beta$ do not depend on $D$.
The best fit in the range $2 \le D \le 1000$,
yields $\alpha=2.3 \pm 0.3$ and $\beta=1.3 \pm 0.2$.  Thus, this example
suggests that the dimensionality of the points does not affect the
performance of the method significantly.

A more careful interpretation is that the method is robust against
irrelevant features present in the characterization of the data.
Clearly, there is only one relevant feature in this problem, which is
given by the projection

\[
   x' =  \frac{\vec y_1 - \vec y_2}{\| \vec y_1 - \vec y_2 \|} 
         \cdot \vec x  \; .
\]

The Bayes classifier, which has the lowest expected error, is
implemented by assigning $\vec x_i$ to cluster 1 if $x'_i < 0$ and to
cluster 2 otherwise.  Therefore we can consider the other $D-1$
dimensions as irrelevant features because they do not carry any
relevant information.  Thus, equation (\ref{eq:Lc}) is telling us
how noise, expressed as the number of irrelevant features present,
affects the performance of the method. 
Adding pure noise variables to the true signal can
lead to considerable confusion when classical methods are used
(Fowlkes {\em et.\ al.\ } 1988).

\subsection{The Iris Data}
\label{sec:iris.data}
The first ``real'' example we present is the time--honored
Anderson--Fisher Iris data, which has become a popular benchmark
problem for clustering procedures. It consists of measurement of four
quantities, performed on each of 150 flowers.  The specimens were
chosen from three species of Iris. The data constitute 150 points in
four-dimensional space.

\begin{figure}[htbp]
  \begin{center}
    \leavevmode
    \psfig{figure=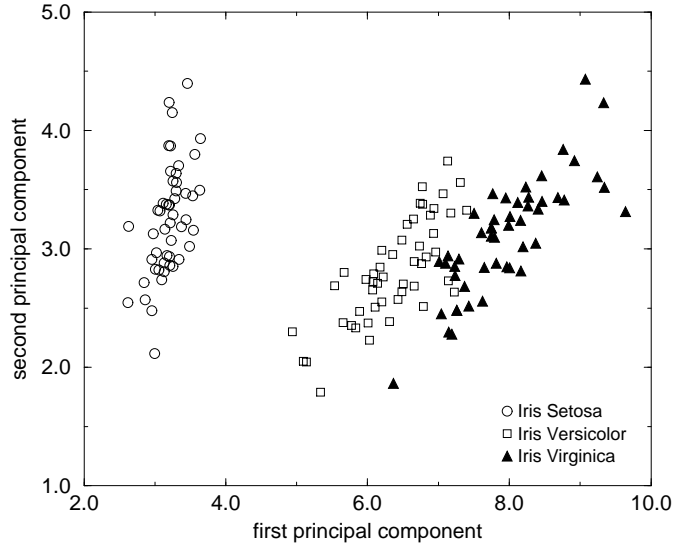,width=10.0cm}
    \caption{Projection of the iris data on the plane spanned by its two
      principal components.}
    \label{fig:iris_PCA}
  \end{center}
\end{figure}

The purpose of this experiment is to present a slightly more
complicated scenario than that of fig. \ref{fig:aros}.  From the
projection on the plane spanned by the first two principal components,
presented on fig. \ref{fig:iris_PCA}, we observe that there is a well
separated cluster (corresponding to the Iris Setosa species) while
clusters corresponding to the Iris Virginia and Iris Versicolor do
overlap.

\begin{figure}[htbp]
  \begin{center}
    \leavevmode
    \psfig{figure=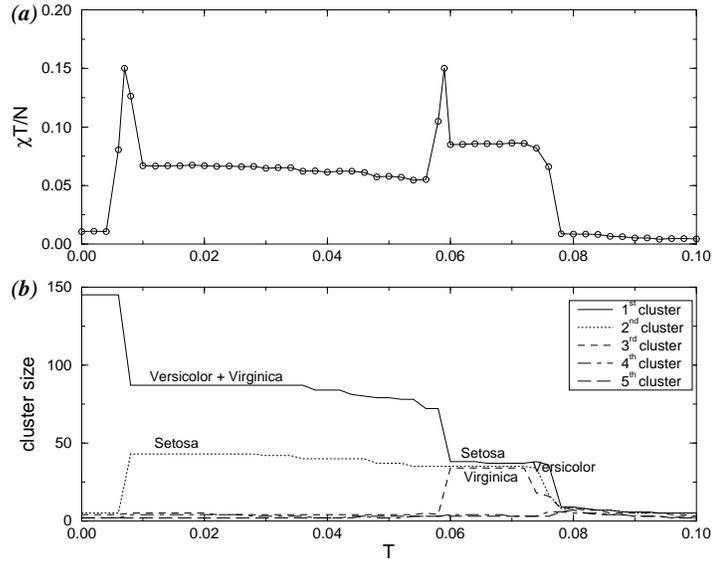,width=10.0cm}
    \caption{$(a)$ The susceptibility density $\frac{\chi T}{N}$ as a 
      function of the temperature and $(b)$ the size of the four biggest clusters
obtained at each temperature for the Iris data.}
    \label{fig:iris_SP}
  \end{center}
\end{figure}

We determined neighbors in the $D=4$ dimensional space according to
the mutual K (K=5) nearest neighbors definition; applied the SPC
method and obtained the susceptibility curve of
Fig.~\ref{fig:iris_SP}$(a)$; it clearly shows {\it two} peaks!  When
heated, the system first breaks into two clusters at $T\approx 0.1$.
At $T_{clus}=0.2$ we obtain two clusters, of sizes 80 and 40; points
of the smaller cluster correspond to the species {\it Iris Setosa}.
At $T\approx 0.6$ another transition occurs, where the larger cluster
splits to two. At $T_{clus}=0.7$ we identified clusters of sizes 45,
40 and 38, corresponding to the species {\it Iris
  Versicolor,Virginica} and {\it Setosa} respectively.

As opposed to the toy problems, the Iris data breaks into clusters in
{\it two stages}. This reflects the fact that two of the three species
are ``closer'' to each other than to the third one; the
SPC method clearly handles very well such hierarchical
organization of the data.  125 samples were classified correctly (as
compared with manual classification); 25 were left unclassified.  No
further breaking of clusters was observed; all three disorder at
$T_{ps}\approx 0.8$ (since all three are of about the same density).

Among all the clustering algorithms used in this work, the minimal
spanning tree procedure obtained the most accurate result, followed by
our method, while the remaining clustering
techniques failed to provide a satisfactory result.

\begin{table}[htbp]
  \begin{center}
    \leavevmode
    \begin{tabular}{| l | r | r | r | }
      \hline
      \hline
      Method                  & biggest & middle  & smallest \\
                              & cluster & cluster & cluster  \\
      \hline
        Minimal Spanning Tree       &   50   &   50  &    50 \\
        Super--Paramagnetic         &   45   &   40  &    38 \\
        Valley Seeking              &   67   &   42  &    37 \\ 
        Complete--Link            &   81   &   39  &    30 \\ 
        Directed Graph              &   90   &   30  &    30 \\ 
        K-Shared Neighbors          &   90   &   30  &    30 \\     
        Single--Link              &  101   &   30  &    19 \\     
        Mutual Neighborhood Value   &  101   &   30  &    19   \\
       \hline
       \hline
    \end{tabular}
    \caption{Best partition obtained with each of the clustering methods.
      Only the minimal spanning tree and the super--paramagnetic
      method returned clusters where points belonging to different
      Iris species were not mixed}
    \label{tab: iris_all}
  \end{center}
\end{table}

\newpage
\subsection{LANDSAT data}
\label{sec:landsat.data}
Clustering techniques have been very popular in remote sensing
applications (Faber {\em et.\ al.\ } 1994, Kelly and White 1993, Kamata
and Kawaguchi 1995; Larch 1994, Kamata  {\em et.\ al.\ } 1991).
Multi-spectral scanners on LANDSAT
satellites sense the electromagnetic energy of the light reflected by
the earth's surface in several bands (or wavelengths) of the spectrum.
A pixel represents the smallest area on earth's surface that can
be separated from the neighboring areas. The pixel size and the number
of bands varies, depending on the scanner; in this case four bands are
utilized, whose pixel resolution is of $80 \times 80$ meters. Two of the
wavelengths are in the visible region, corresponding approximately to
green (0.52 to 0.60 $\mu_m$) and red (0.63 to 0.69$\mu_m$) and the
other two are in the near--infrared (0.76 to 0.90 $\mu_m$) and
mid--infrared (1.55 to 1.75 $\mu_m$) regions.  The wavelength interval
associated with each band is tuned to a particular cover category. For
example the green band is useful for identifying areas of shallow
water, such as shoals and reefs, whereas the red band emphasizes urban
areas.

The data consist of 6437 samples that are contained in a rectangle of
$82 \times 100$ pixels.  Each ``data point'' is described by 36 features
that correspond to a $3 \times 3$ square of pixels.
A classification label (ground truth) of the central pixel is also
provided.  The data is given in random order and certain samples have
been removed, so that one cannot reconstruct the original image.
The data was provided by Srinivasan (1994) and is available at the UCI
Machine Learning Repository (Murphy and Aha 1994).

The goal is to find the ``natural classes'' present in the data
(without using its labels, of course).  The quality of our results is
determined by the extent to which the clustering reflects the six terrain
classes present in
the data: red soil, cotton crop, grey soil, damp grey soil, soil with
vegetation stubble and very damp grey soil.  This exercise is close to
a real problem of remote sensing, where the true labels (ground truth)
on the pixels is not available, and therefore clustering techniques
are needed to group pixels on the basis of the sensed observations.

\begin{figure}[htbp]
  \begin{center}
    \leavevmode
    \psfig{figure=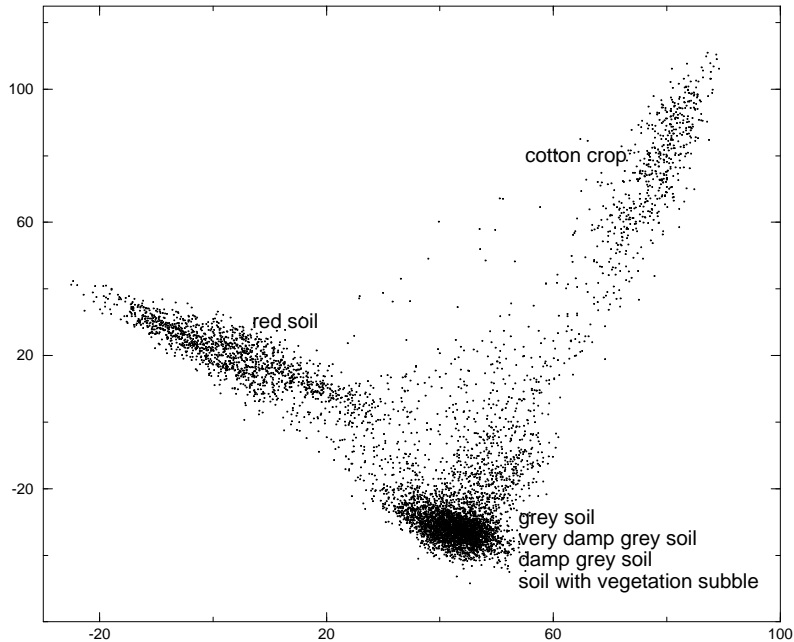,width=13.0cm}
    \caption{Best two dimensional projection pursuit among the first six
      solutions for the LANDSAT data}
    \label{fig:landsat_pp}
  \end{center}
\end{figure}
We used the projection pursuit method (Friedman 1987), which is a
dimension--reducing transformation, in oder to gain some knowledge
about the organization of the data.  Among the first six two--dimensional
projections that were produced we present in figure
\ref{fig:landsat_pp} that one which reflects best the (known)
structure of the data.
We observe the that:
(a) the clusters differ in their density, 
(b) there is unequal coupling between clusters, and
(c) the density of the points within a cluster
is not uniform; it decreases towards the perimeter of the cluster.

\begin{figure}[htbp]
  \begin{center}
    \leavevmode
    \psfig{figure=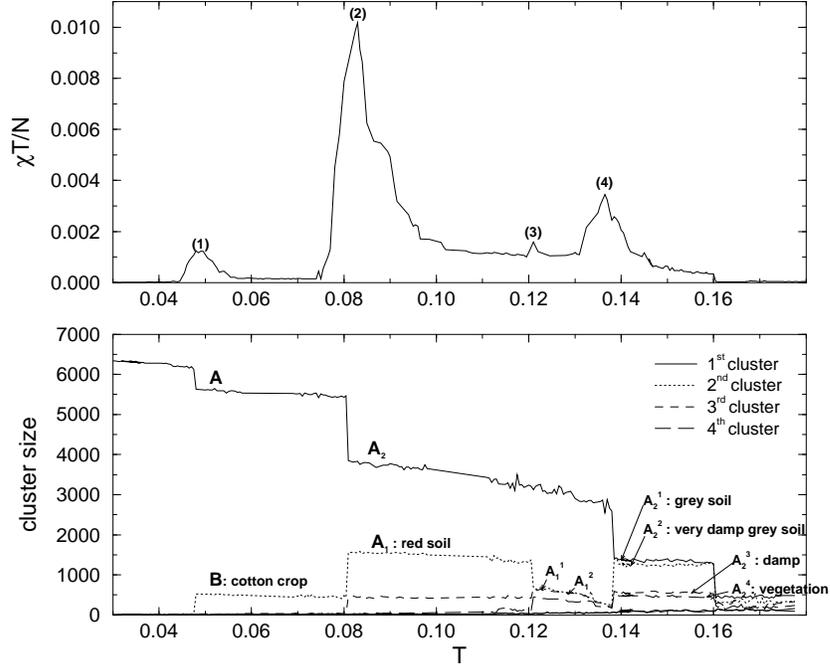,width=12.0cm}
    \caption{$(a)$ Susceptibility density $\frac{\chi T}{N}$ of the Landsat
      data as a function of the temperature T. The number in
      parenthesis indicate the phase transitions.  $(b)$ The sizes of
      the four biggest clusters at each temperature. The jumps
      indicate that a cluster has been split. Symbols $A$,$B$,$A_i$
      and $A_i^j$ corresponds to the hierarchy depicted in 
      fig. \protect{\ref{fig:landsat_hier}}}.
    \label{fig:landsat_sp}
  \end{center}
\end{figure}
\begin{figure}[htbp]
  \begin{center}
    \leavevmode
    \psfig{figure=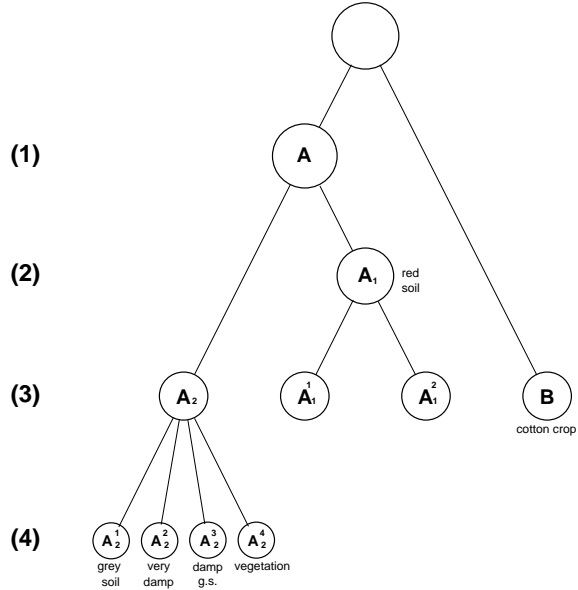,width=7.50cm}
    \caption{The Landsat data structure reveals a hierarchical structure.
    The number in parenthesis corresponds to the phase transitions
    indicated by a peak in the
    susceptibility (fig.~\protect{\ref{fig:landsat_sp}})}
    \label{fig:landsat_hier}
  \end{center}
\end{figure}

The susceptibility curve Fig.~\ref{fig:landsat_sp}$(a)$ reveals {\it
four} transitions, that reflect the presence of the following
hierarchy of clusters (see fig.~\ref{fig:landsat_hier}).
At the lowest temperature two clusters $A$ and
$B$ appear. Cluster $A$ splits at the second transition into $A_1$ and
$A_2$. At the next transition cluster $A_1$ splits into $A_1^1$ and
$A_1^2$.  At the last transition cluster $A_2$ splits into four
clusters $A_2^i, i=1...4$. At this temperature the clusters $A_2$ and
$B$ are no longer identifiable; their spins are in a disordered state,
since the density of points in $A_2$ and $B$ is significantly smaller
than within the $A_1^i$ clusters.  Thus the super--paramagnetic method
overcomes the difficulty of dealing with clusters of different
densities by analyzing the data at several temperatures.
This hierarchy indeed reflects the structure of the data. Clusters
obtained in the range of temperature 0.08 to 0.12 coincides with the
picture obtained by projection pursuit; cluster $B$ corresponds to 
cotton crop terrain
class, $A_1$ to red soil and the remaining four terrain classes are
grouped in the cluster $A_2$. The clusters $A_1^1$ and $A_1^2$ are a
partition of the red soil\footnote{This partition of the red soil is 
{\em not} reflected in the ``true'' labels. It would be of interest
to reevaluate the labeling and try to identify the features that
differentiate the two categories of red soil that were discovered by our
method.}, while $A_2^1$, $A_2^2$, $A_2^3$ and $A_2^4$
correspond, respectively, to the classes grey soil, very damp grey
soil, damp grey soil and soil with vegetation stubble.
97\% purity
was obtained, meaning that points belonging to different categories
were almost never assigned to the same cluster.

Only the optimal answer of Fukunaga's valley seeking, and our SPC
method succeeded in recovering the structure of the LANDSAT data.
Fukunaga's method, however, yielded for different (random) initial
conditions grossly different answers, while our answer was stable.

\subsection{Isolated Letter Speech Recognition}
\label{sec:isolet.data}
In the isolated--letter speech recognition task, the ``name'' of a
single letter is pronounced by a speaker.  The resulting audio signal
is recorded for all letters of the English alphabet for many speakers.
The task is to find the structure of the data, which is expected to be
a hierarchy reflecting the similarity that exists between different
groups of letters, such as $\{ B, D \}$ or $\{M, N \}$ which differ
only in a single articulatory feature.  This analysis could be useful,
for instance, to determine to what extent the chosen features succeed
in differentiating the spoken letters.

We used the ISOLET database of 7797 examples created by Ron Cole
(Fanty and Cole 1991) which is available at the UCI
machine learning repository (Murphy and Aha 1994).
The data was recorded from 150
speakers balanced for sex and representing many different accents and
English dialects.  Each speaker pronounce each of the 26 letters twice
(there are 3 examples missing).  Cole's group has developed a set of
617 features describing each example.  All attributes are continuous
and scaled into the range $-1$ to $1$.  The features include spectral
coefficients, contour features, sonorant, pre--sonorant, and
post--sonorant features.  The order of appearance of the features is
not known.

\begin{figure}[htbp]
  \begin{center}
    \leavevmode
    \centerline{
        \psfig{figure=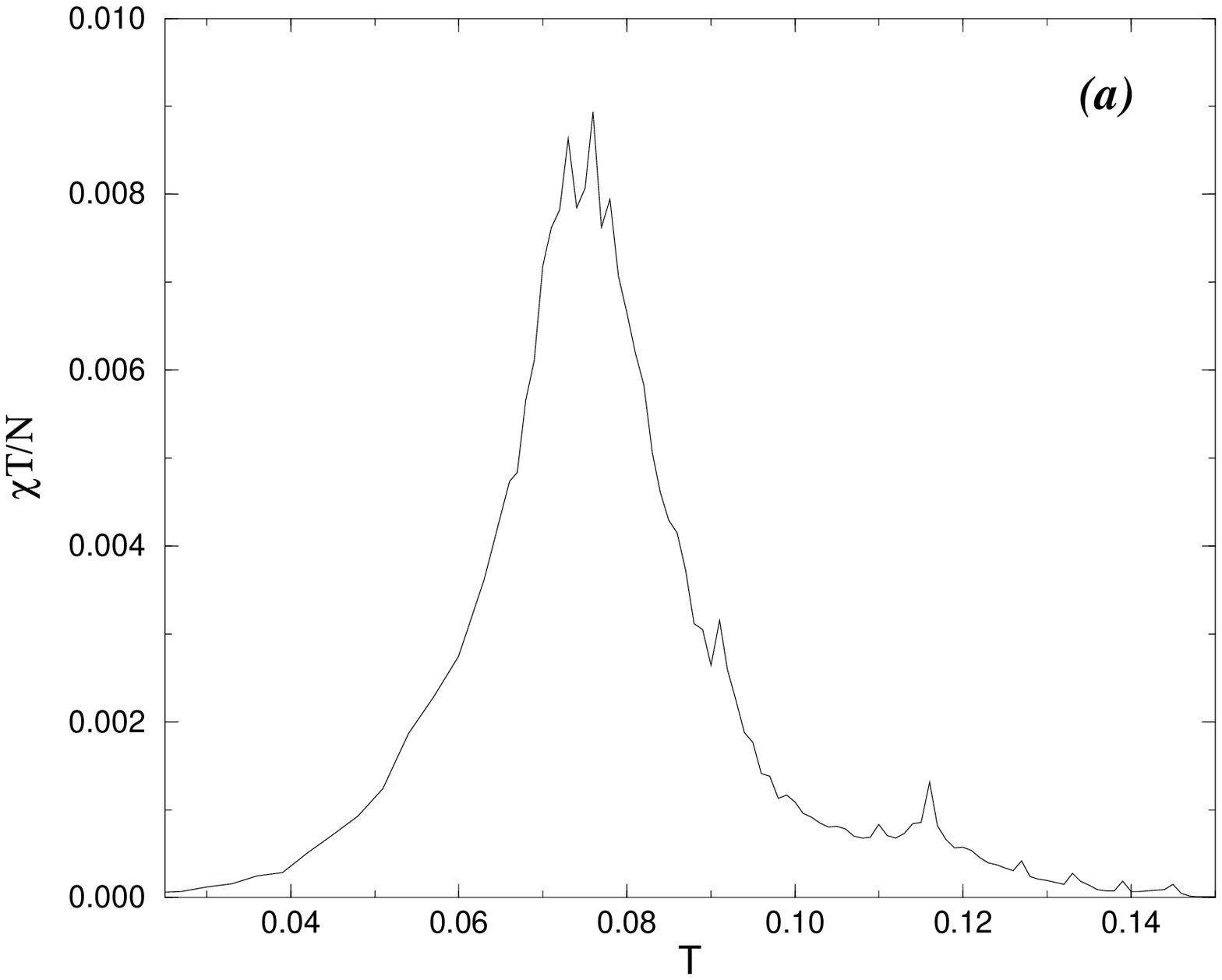,height=10.0cm}
        }
      \centerline{
        \psfig{figure=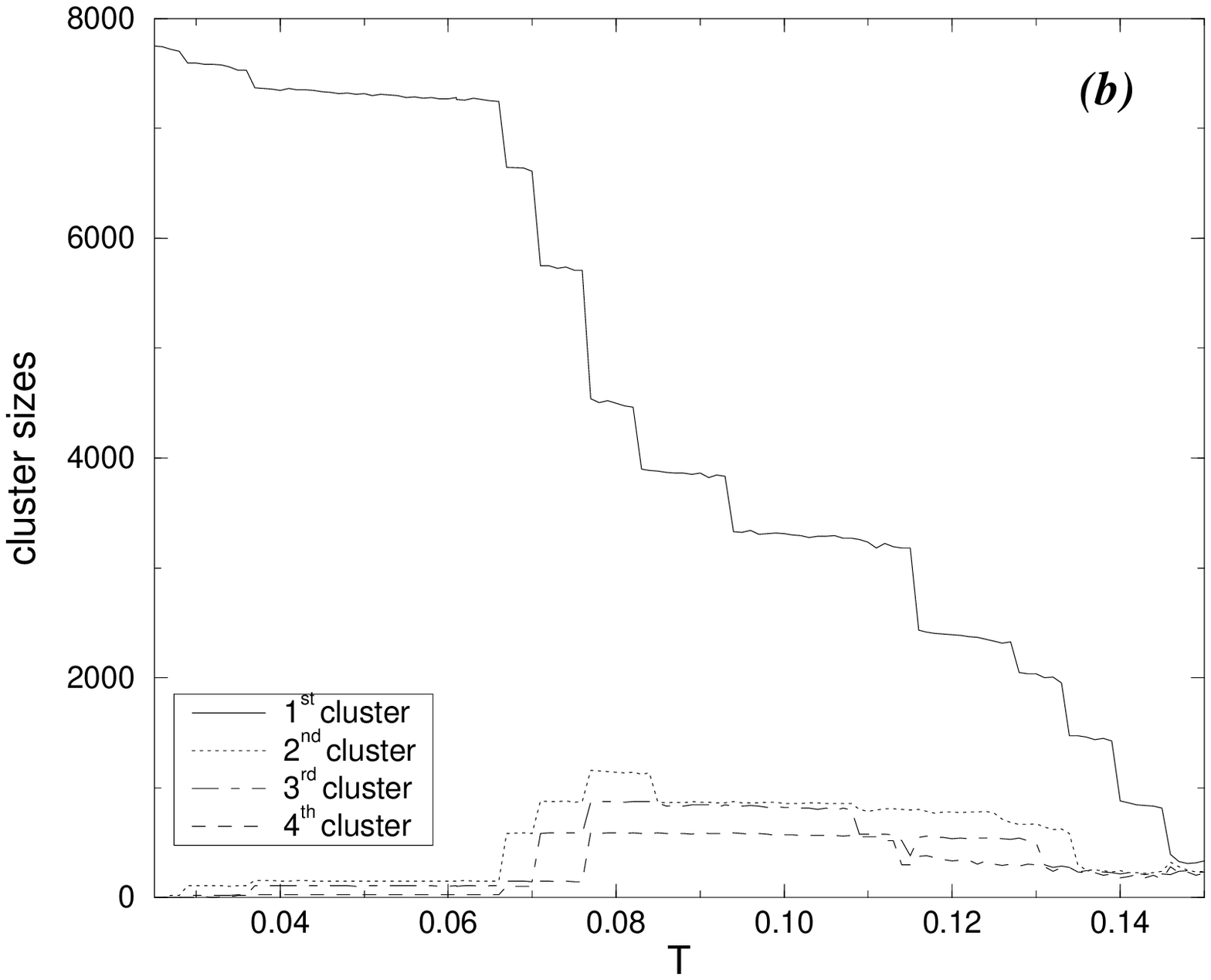,height=10.0cm}
        }
    \caption{$(a)$ Susceptibility density as a function of the temperature
      for the isolated--letter speech--recognition data. $(b)$ Size of the
      four biggest clusters returned by the algorithm for each temperature.}
    \label{fig:isolet_SP}
  \end{center}
\end{figure}

\begin{figure}[htbp]
  \begin{center}
    \leavevmode
    \centerline{
      \psfig{figure=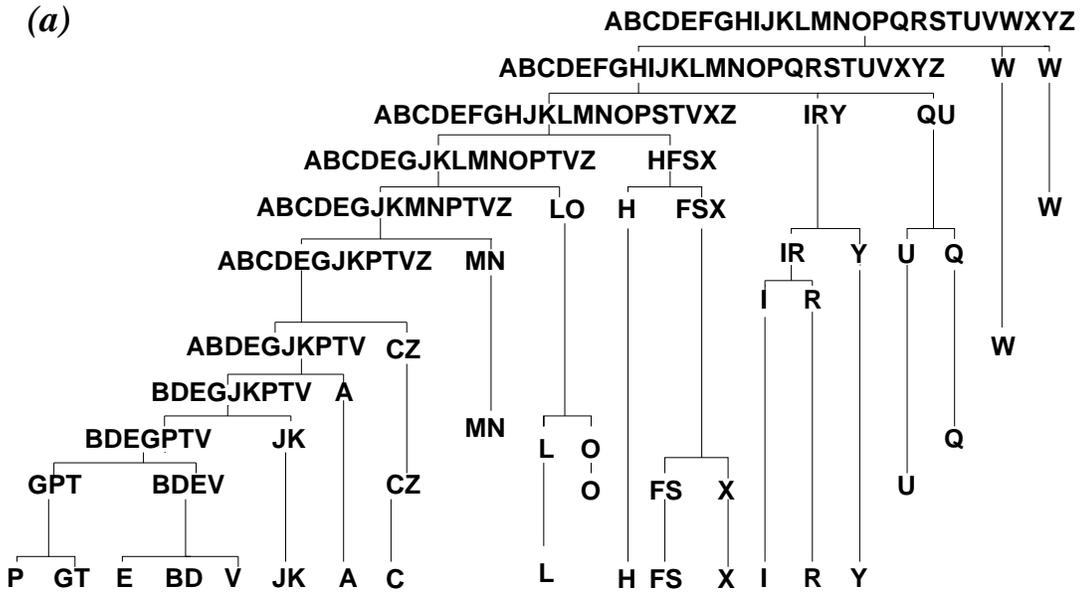,width=15.0cm} 
      }
    \vskip \baselineskip
    \centerline{
      \psfig{figure=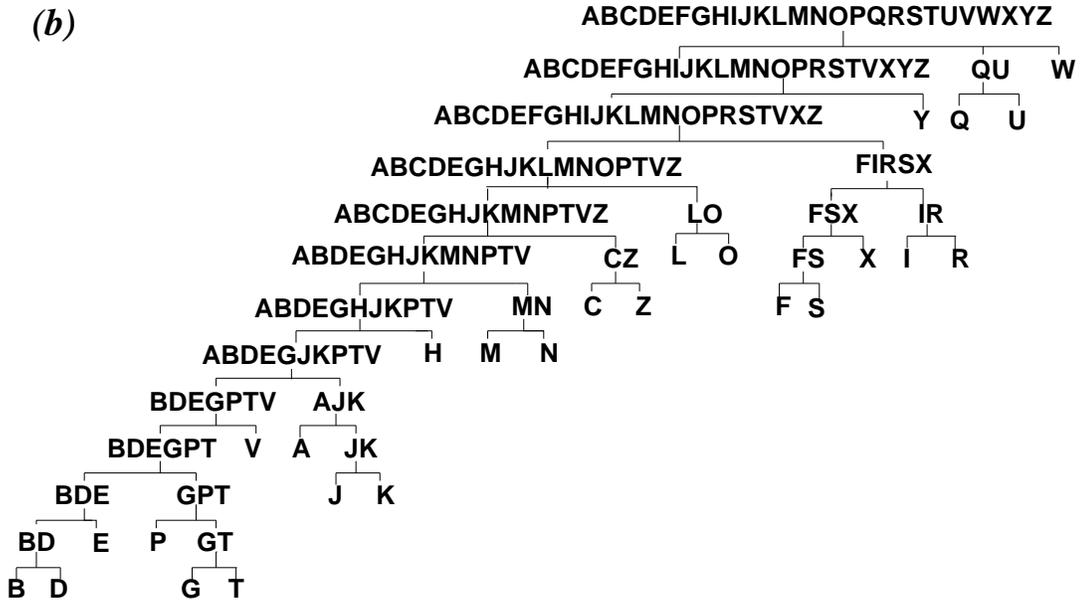,width=15.0cm}
      }
    \caption{Isolated--letter speech--recognition hierarchy obtained by $(a)$
      the Super--Paramagnetic method and $(b)$ using the labels of the data and
      assuming each letter is well represented by a center.}
    \label{fig:isolet_hier}
  \end{center}
\end{figure}

We applied the SPC method and obtained the susceptibility curve shown
in figure \ref{fig:isolet_SP}$(a)$ and the cluster size versus temperature
curve presented in fig \ref{fig:isolet_SP}$(b)$.  The resulting partitioning
obtained at different temperatures can be cast in hierarchical form,
as presented in fig. \ref{fig:isolet_hier}$(a)$.

We also tried the projection pursuit
method; but none of the first six 2-dimensional projections
succeeded to reveal any relevant 
characteristic about the structure of the data. 
In assessing the extent to which the SPC method succeeded to recover the
structure of the data, we built a ``true'' hierarchy
by using the known labels of the examples.
To do this, we first calculate the center of each class (letter)
by averaging over all the examples belonging to it. Then a matrix
$26 \times 26$ of
the distances between these centers is constructed.
Finally, we apply the single--link method to construct a hierarchy,
using this proximity matrix.
The result is presented in figure \ref{fig:isolet_hier}$(b)$.
The purity of the clustering was again very high ($93 \%$); 
and $35 \%$ of the samples were left as  unclassified points.  
The CPCC validation index 
(Jain and Dubes 1988) is equal to 0.98 for this graph, which indicates 
that this hierarchy fits very well the data. Since our method does not have
a natural length scale defined at each resolution, we cannot 
use this index for our tree. Nevertheless,
 the good quality of our tree, presented in 
figure \ref{fig:isolet_hier}$(a)$, is indicated by
the good agreement between it and the tree of Fig. \ref{fig:isolet_hier}$(b)$.
Needless to say, in order to construct the ``reference'' tree depicted in 
fig.\ref{fig:isolet_hier}$(b)$ , the correct
label of each point must be known.

\sction{Complexity and computational overhead}
Non-parametric  clustering is performed in two main stages:
\begin{description}

\item[\bf Stage 1:] Determination of the ``geometrical structure''
  of the problem. 
  Basically a number of nearest neighbors of each point has to be found,
  using any reasonable algorithm, such as identifying
  the points lying inside a sphere of a given radius, or a given number
  of closest neighbors (like in the SPC algorithm).
  
\item[\bf Stage 2:] Manipulation of the data. Each method is characterized
by a specific processing of the data.

\end{description}

For almost all methods, including SPC,  complexity is
determined by the first stage because it deserves more computational
effort than the data manipulation itself. Finding the nearest neighbors is an 
expensive task, the complexity of branch and bound algorithms 
(Kamgar-Parsi and Kanal 1985) is 
of order ${\cal O}(N^\nu \log N)$ ($1<\nu<2$).
Since this operation is {\em common for all} non--parametric 
clustering methods, 
any extra computational overhead our algorithm may have over 
some other non--parametric 
method must be due to the difference between the costs of the
 manipulations
performed beyond this stage.
The second stage, in the SPC method,
consists of equilibrating a system at each temperature.
In general, the complexity is of order $N$(Binder and Heermann 1988,
Gould and Tobochnik 1988).

{\bf Scaling with N:} 
The main reason for choosing an unfrustrated ferromagnetic system, 
versus to a spin--glass (where negative interactions are allowed),
is that ferromagnets reach thermal equilibrium very fast. 
Very efficient Monte Carlo algorithms 
(Wang and Swendsen 1990, Wolf 1989, Kandel and Domany 1991)
were developed for these systems, in which
the number of sweeps needed for thermal equilibration 
is small at the temperatures of interest.
The number of operations required for each Swendsen--Wang 
Monte Carlo sweep scales linearly with the number of edges;
{\em i.e.} it is of order $K \times N$ (Hoshen and Kopelman 1976). 
In all the examples of this article we used a fixed number of sweeps 
($M = 1000$). 
Therefore, the fact that the SPC method relies on a 
stochastic algorithm does not prevent it from being efficient.

{\bf Scaling with D:} The equilibration stage does not depend on the
dimension of the data, $D$. In fact it is not necessary to know the 
dimensionality of the data as long as the distances between neighboring 
points are known. 

Since the complexity of the equilibration stage is of order $N$ and
does not scale with $D$,  the complexity of the method is determined
by the search for the nearest neighbors. Therefore,
we conclude that {\em the complexity of our method does not exceed that
of the most efficient deterministic non--parametric algorithms}.

For the sake of concreteness, we present the running times, corresponding to
the second stage, on an HP--9000 (series K200) machine for two problems: 
the LANDSAT (sec. \ref{sec:landsat.data})
and ISOLET data (sec. \ref{sec:isolet.data}).
The corresponding running times were $1.97$ and 
$2.53$ minutes per temperature respectively 
(0.12 and 0.15 sec per sweep per temperature). 
Note that there is a good agreement with the discussion presented above;
the ratio of the CPU times is close to the ratio
of the corresponding total number of edges (18,388 in the LANDSAT and
22,471 in the ISOLET data set), 
and there is no dependence on the dimensionality .
Typical runs involve about 20 temperatures which leads to
40 and 50 minutes of CPU. This number of temperatures can be significantly
reduced by using the Monte Carlo histogram method (Swendsen 1993)
where a set of simulations at small number of temperatures suffices to
calculate thermodynamic averages for the complete temperature range 
of interest.
Of all the deterministic methods we used, the most efficient one 
is the minimal spanning tree. Once the tree is built,
it requires only 19 and 23 seconds of CPU 
respectively for each set of clustering parameters. However, the actual 
running time is determined by how long one spends searching for the 
optimal parameters in the  (3--dimensional) 
parameter space of the method.
The other non--parametric methods presented in this paper were not
optimized and therefore comparison of their running times could be
misleading. 
For instance, we used Johnson's algorithm for implementing the single and 
complete linkage which requires ${\cal O}(N^3)$ operations for
recovering all the hierarchy, but faster versions, based on Minimal 
Spanning trees require less operations.
Running Friedman's projection pursuit algorithm\footnote{
  We thank Jerome Friedman for allowing public use his program.
},
whose results are presented in 
fig.\ \ref{fig:landsat_pp}, required 55 CPU minutes for LANDSAT. 
For the case of the ISOLET data (where $D=617$) the difference was dramatic; 
projection pursuit required {\it more than a week} of CPU time,
while SPC required about one hour. The reason is that our 
algorithm does not scale with the dimension of the data $D$,
whereas the  complexity of projection pursuit increases very fast 
with $D$.

\sction{Discussion}
This work proposes a new approach to non-parametric clustering, based
on a physical, magnetic analogy.  The mapping onto the magnetic problem is
very
simple; a Potts spin is assigned to each data point, and short-range 
ferromagnetic
interactions between spins are introduced.  The strength of these
interactions decreases with distance.  The thermodynamic system
defined in this way presents different self organizing regimes and the
parameter which determines the behavior of the system is the temperature.
As the temperature is varied the system undergoes many phase transitions.
The idea is that each phase reflects a particular data structure
related to a particular length scale of the problem.  Basically, the
clustering obtained at one temperature that belongs to a 
specific phase should not differ substantially
from the partition obtained at another temperature in the same
phase. On the other hand the clustering obtained at two temperatures
corresponding to different phases must be significantly different, reflecting
different organization of the data.  These ordering
properties are reflected in the susceptibility $\chi$ and the
spin--spin correlation function $G_{ij}$.  The susceptibility turns
out to be very useful for signaling the transition between 
different phases of the system.  The correlation function $G_{ij}$ is
used as a similarity index, whose value is not determined only by the
distance between sites $v_i$ and $v_j$, but also by the density of
points near and between these sites.
Separation of the
spin--spin correlations $G_{ij}$ into strong
and weak, as evident in fig.  \ref{fig:aros_hist}$(a)$, reflects the
existence of two categories of collective behavior.
In contrast, as shown in figure
\ref{fig:aros_hist}$(b)$, the frequency distribution of distances $d_{ij}$
between neighboring points of Fig.~\ref{fig:aros}  does not even hint
that a natural cut-off distance, which separates neighboring points
into two categories, exists.  Since the double
peaked shape of the correlations' distribution persists at all relevant
temperatures, the separation into strong and weak correlations is a
robust property of the proposed Potts model. 

This procedure is stochastic, since we use a Monte Carlo procedure to
``measure'' the different properties of the system, but it is completely
insensitive to  initial conditions. Moreover, the cluster distribution
as a function of the temperature is known. Basically, there is a competition
between the positive interaction which encourages the spins to be aligned
(the energy, which appears in the exponential of the Boltzmann weight,
is minimal when all points belong to a single cluster) and the 
thermal disorder (that assigns a ``bonus'' that grows exponentially 
with the number of uncorrelated spins and, hence, with 
the number of clusters).

We have shown that this method is robust in the presence of noise, and that
it is able to recover hierarchical structure of the data without
enforcing the presence of clusters.
We have also confirmed that the super--paramagnetic method is successful 
in real life problems, where existing methods failed to overcome the
difficulties posed
by the existence of different density distributions and many
characteristic lengths in the data.

Finally we wish to re-emphasize the  
aspect we view as the main advantage of our method: it's 
{\it generic applicability}. It is likely and natural to expect that for just 
about any underlying distribution of data one will be able to find a 
particular method, tailor-made to handle the particular distribution, whose
performance will be better than that of SPC. If however, there is no advance
knowledge of this distribution, one cannot know which of the existing methods 
fits best and should be trusted. SPC, on the other hand, will find any 
"lumpiness" (if it exists) of the underlying data, without any
fine-tuning of its parameters.

\section*{Acknowledgments \hrulefill} 
We thank I. Kanter for many useful discussions.
This research has been supported by the Germany-Israel Science
Foundation (GIF).

\section*{Appendix: Clusters and the Potts model \hrulefill}
\label{sec:Cluster.and.Potts}
The Potts model can be mapped
onto a random--cluster problem (Fortuin and Kasteleyn, 1972, Coniglio
and Klein 1980, Edwards and Sokal 1988). In this formulation clusters
are defined as connected graph components governed by a specific
probability distribution. We present here this alternative formulation
in order to give another motivation for the
super--paramagnetic method as well as to facilitate its comparison to
graph based clustering techniques.

Consider the following graph based model whose basic entities are bond
variables $n_{ij}=0,1$ residing on the edges $<i,j>$ connecting neighboring
sites $v_i$ and $v_j$. When $n_{ij} = 1$ the bond between sites $v_i$
and $v_j$ is ``occupied'', and when $n_{ij} = 0$ the bond is
``vacant''.  Given a configuration ${\cal N} = \left \{ n_{ij} \right
\}$, random--clusters are defined as the vertices of the connected
components of the occupied bonds (where a vertex connected to
``vacant'' bonds only is considered a cluster containing a single point). The
random cluster model is defined by the probability distribution
\begin{equation} \label{eq:Fortuin} 
     W \! {\scriptstyle \left ( {\cal N} \right )}
     \; = \; 
     \frac{ q^{C {\scriptscriptstyle \left ({\cal N} \right )}}}{Z} \;
        \prod_{<i,j>}
         p_{ij}^{n_{ij}} \;  \left ( 1-p_{ij} \right )^{( 1-n_{ij})} \; ,
\end{equation}
where $C {\scriptstyle \left ( {\cal N} \right )}$ is the number of
clusters of the given bond configuration, the partition sum $Z$ is
a normalization constant, and the parameters $p_{ij}$ fulfill 
$1\geq p_{ij} \geq 0$. 

The case $q=1$ is the percolation model where the joint probability
(\ref{eq:Fortuin}) factorizes into a product of independent factors
for each $n_{ij}$. Thus the state of each bond is independent of the
state of any other bond. This implies for example that the most
probable state is found simply by setting $n_{ij}=1$ if $p_{ij}>0.5$
and $n_{ij}=0$ otherwise.  By choosing $q>1$ the weight of any bond
configuration ${\cal N}$ is no longer the product of local independent
factors.  Instead the weight of a configuration is also influenced by
the spatial distribution of the occupied bonds, since configurations
with more random--clusters are given a higher weight.  For instance it
may happen that a bond $n_{ij}$ is likely to be vacant while a bond
$n_{kl}$ is likely to be occupied even though $p_{ij}=p_{kl}$.  This
can occur if the vacancy of $n_{ij}$ enhances the number of
random--clusters, while sites $v_k$ and $v_l$ are connected through
other (than $n_{kl}$) occupied bonds.

Surprisingly there is a deep connection between the random--cluster model
and the seemingly unrelated Potts model. The basis for this connection 
(Edwards and Sokal 1988) is a
joint probability distribution of Potts spins and bond variables:

\begin{equation}
   P \! {\scriptstyle{ ({\cal S},{\cal N}) }} =
   \frac{1}{Z} \prod_{<i,j>} \left [ 
     (1-p_{ij}) \; (1-n_{ij}) \; + \; p_{ij} \; n_{ij} \; \delta_{s_i,s_j}
   \right ] \; .
   \label{eq: joint P}
 \end{equation}
The marginal probability $W \! {\scriptstyle \left ( {\cal N} \right )}$ is
obtained by summing $P \! {\scriptstyle{({\cal S},{\cal N})}}$ over all Potts
spin configurations. On the other hand by setting
\begin{equation} \label{eq:Pij}
    p_{ij} = 1 - \exp{ \!\! \left ( - \frac{J_{ij}}{T} \right )} \; , 
\end{equation}
and summing $P \; {\scriptstyle { ({\cal S},{\cal N})}}$
over all bond configurations the marginal 
probability (\ref{eq:Boltzmann}) is obtained.

The mapping between the Potts spin model and the random--cluster model implies
that the super--paramagnetic clustering method can be formulated in terms of
the random--cluster model. One way to see this is to realize that
the SW--clusters {\em are} actually the random--clusters. That is
the prescription given in Sec. \ref{sec:SWmethod} for generating the
SW--clusters is defined through the conditional probability
$
  P \! {\scriptstyle{({\cal N} | {\cal S})}} = 
  P \! {\scriptstyle{({\cal S}} , {\cal N})} / P \! {\scriptstyle{({\cal S})}}
$.  Therefore by
sampling the spin configurations obtained in the Monte Carlo sequence
(according to probability $P \! {\scriptstyle{({\cal S})}}$), the bond
configurations obtained are generated with probability                     
$W \!{\scriptstyle \left ( {\cal N} \right )}$. In addition, remember that
 the Potts
 spin--spin correlation function $G_{ij}$ is measured by using equation
 (\ref{eq:correlation}) and relying on the statistics of the SW--clusters. 
Since the clusters are obtained through the spin--spin correlations they
can be determined directly from the random--cluster model.

One of the most salient features of the super--paramagnetic method is its 
probabilistic approach as opposed to the deterministic one taken
in other methods.  Such deterministic schemes can indeed be recovered
in the zero temperature limit of this formulation 
(equations (\ref{eq:Fortuin}) and  (\ref{eq:Pij}) ); 
at $T=0$ only the bond
configuration ${\cal N}_0$ corresponding to the ground state, appears
with non-vanishing probability.
Some of the existing clustering methods can be formulated as
deterministic--percolation models ($T=0$, $q=1$). For instance, the
percolation method proposed by Dekel and West (1985) is obtained
by choosing the coupling between spins 
$J_{ij} = \theta \! \left (R - d_{ij} \right)$;
that is, the interaction between spins $s_i$ and $s_j$ is equal to
one if its separation is smaller than the clustering parameter $R$ 
and zero otherwise.
Moreover, the
single--link hierarchy (see for example Jain and Dubes, 1988) is
obtained by varying the clustering parameter $R$. Clearly, in these
processes the reward on the number of clusters is ruled out and
therefore only pairwise information is used in those procedures.

Jardine and Sibson (1971) attempted to list the essential
characteristics of useful clustering methods and concluded that the
single--link method was the only one that satisfied all the
mathematical criteria.  However, in practice it performs poorly
because single--link clusters easily chain together and are often
``straggly''. Only a single connecting  edge  is needed to
merge two large clusters.  To some extent the super--paramagnetic method
overcomes this problem by introducing a bonus on the number of
clusters which is reflected by the fact that the system prefers to
break apart clusters that are connected by a small number of bonds.

Fukunaga's (1990) {\em valley--seeking} method is recovered
in the case $q > 1$ with interaction between spins  
$J_{ij} = \theta \! \left (R - d_{ij} \right)$. In this case, the Hamiltonian
(\ref{eq:Hamiltonian}) is just the class separability measure of this
algorithm where a Metropolis relaxation at $T=0$ is used to minimize
it.  The relaxation process terminates at some local minimum of the
energy function, and points with the same spin value are assigned to a
cluster. This procedure depends strongly on the initial conditions and
is likely to stop at a metastable state that does not correspond to
the correct answer.

\section*{References \hrulefill}

\begin{description}
  
\item Ahuja,~N.\ (1982).  ``Dot pattern processing using Voronoi
    neighborhood'',~{\em IEEE Transactions on Pattern Analysis and Machine
  Intelligence} {\bf PAMI 4}, 336--343.
 
\item Ball,~G., and Hall,~D. 1967. ``A clustering technique for
  summarizing multivariate data'', {\em Behavioral Science} {\bf 12},
  153--155.

\item Baxter,~R.~J. 1973. ``Potts model at the critical temperature'',
  {\em Journal of Physics} {\bf C 6}, L445--448. 
  
\item Baraldi,~A., and Parmiggiani,~F. 1995.  ``A neural network for
  unsupervised categorization of multivalued input patterns: an
  application to satellite image clustering'', {\em IEEE Transactions
    on Geoscience and Remote Sensing} {\bf 33}(2), 305--316.
  
\item Binder,~K., and Heermann,~D.W. 1988. {\em Monte Carlo Simulations
    in statistical physics, an introduction}. Springer--Verlag, Berlin

\item Billoire,~A., Lacaze,~R., Morel,~A., Gupta,~S.\., Irback,~ A., and
  Petersson,~B.  1991.  ``Dynamics near a first--order phase transition
  with the Metropolis and Swendsen--Wang algorithms.''  {\em Nuclear
    Physics} {\bf B 358}, 231 -- 248.
  
\item Blatt,~M., Wiseman,~S., and Domany,~E. 1996a.
  ``Super--paramagnetic clustering of data'', {\em Physical Review
    Letters} {\bf 76}, 3251--3255.
  
\item Blatt,~M., Wiseman,~S., and Domany,~E. 1996b.  ``Clustering
  data through an analogy to the Potts model'' to appear in {\em
    Advances in Neural Information Processing Systems} {\bf 8},
  Touretzky, Mozer, Hasselmo, eds., MIT Press.
  
\item Blatt,~M., Wiseman,~S., and Domany,~E. 1996c. 
  ``Method and apparatus for clustering data'', 
  USA patent application (pending). 

\item Buhmann,~J.M., and K{\"u}hnel,~H. 1993.  ``Vector
  quantization with complexity costs'', {\em IEEE Transactions
    Information Theory} {\bf 39}, 1133 (1993).
  
\item Chen,~S., Ferrenberg,~ A.M., and Landau~D.P. 1992. 
    ``Randomness--induced second--order transitions in the
    two--dimensional eight--state Potts model: a Monte Carlo study'',
  {\em Physical Review Letters}, {\bf 69} (8),  1213--1215.

\item Coniglio,~A., and Klein,~W. 1981.
 ``Thermal phase transitions at the percolation--threshold'',
 Physics Letters {\bf A 84}, 83--84.
  
\item Cranias,~L., Papageorgiou,~H.', and Piperidis,~S. 1994.
  ``Clustering: a technique for search space reduction in
  example--based machine translation'', proceedings of the {\em 1994
    IEEE International Conference on Systems, Man, and Cybernetics.
    Humans, Information and Technology}, {\bf 1}, 1--6.  IEEE, New
  York.
  
\item Dekel,~A., and West,~M.J. 1985.  ``On percolation as a
  cosmological test'', {\em The Astrophysical Journal} {\bf 288},
  411--417.
  
\item Duda,~R.O., and Hart,~P.E. 1973. {\em Pattern Classification and
    Scene Analysis}.  Wiley--Interscience, New York.

\item Edwards,~R.G., and Sokal,~A.D. 1988.
``Generalization of the Fortuin--Kasteleyn--Swendsen--Wang representation
and Monte Carlo algorithm'', {\em Physical review D} {\bf 38}, 2009--2012.

\item Faber,~V., Hochberg,~J.G., Kelly,~P.M., Thomas,~T.R., and
  White,~J.M. 1994.  ``Concept extraction, a data--mining technique''.
  {\em Los Alamos Science}, {\bf 22}, 122--149.

\item Fanty,~M., and Cole,~R. 1991. ``Spoken letter recognition'',
{\em Advances in Neural Information Processing Systems} {\bf 3}, 
220--226.
Lippmann, Moody and Touretzky, eds., Morgan--Kaufmann, San Mateo.

\item Foote,~J.T., and Silverman,~H.F. 1994. ``A model distance
  measure for talker clustering and identification'', proceedings of
  the {\em 994 IEEE International Conference on Acoustics, Speech and
    Signal Processing} {\bf 1}, 317--320.  IEEE, New York.

\item Fowlkes,~E.B., Gnanadesikan,~R., and Kettering,~J.R. 1988.
  ``Variable selection in clustering'', {\em Journal of
    classification}, {\bf 5} 205--228.

\item Friedman,~J.H. 1987. ``Exploratory projection pursuit'', {\em
    Journal of the American statistical association} {\bf 82},
  249--266.

\item Fu,~Y., and Anderson,~P.W. 1986. ``Applications of statistical 
mechanics to NP--complete problems in combinatorial optimization''. 
 {\em Journal of Physics A: Math. Gen.} {\bf 19} 1605--1620.
  
\item Fukunaga,~K. 1990. {\em Introduction to statistical Pattern
    Recognition}. Academic Press, San Diego.

\item Fortuin,~C.M., and Kasteleyn,~P.W. 1972. ``On the random--cluster 
model'', {\em Physica} (Utrecht), {\bf 57}, 536--564.

\item Gdalyahu,~Y.,  Weinshall,~D., unpublished.

\item Gould,~H., and Tobochnik,~J. 1989. {\em An introduction to
    computer simulation methods}, part II. Addison--Wesley, New York.

\item Gould,~H., and Tobochnik,~J. 1989.  ``overcoming critical
  slowing down'', {\em Computers in physics} 82--86.

\item Hennecke,~M, and Heyken,~U. 1993.
``Critical--dynamics of cluster algorithms in the diluite Ising--model'',
{\em Journal of Statistical Physics} {\bf 72}, 829--844.

\item Hertz,~J., Krogh,~ A., and Palmer,~R. 1991.  {\em Introduction to
    the theory of neural computation} Addison--Wesley, Redwood City.

\item Hoshen,~J. and Kopelman,~R. 1976. ``Percolation and cluster 
distribution. I. Cluster multiple labeling technique and critical 
concentration algorithm'' {\em Physical Review} {\bf B 14},
3438--3445.
  
\item Iokibe,~T. 1994.  ``A method for automatic rule and membership
  function generation by discretionary fuzzy performance function and
  its application to a practical system'', Proceedings of the {\em
    First International Joint Conference of the North American Fuzzy
    Information Processing Society Biannual Conference.  The
    Industrial Fuzzy Control and Intelligent Systems Conference, and
    the NASA Joint Technology Workshop on Neural Networks and Fuzzy
    Logic.}, 363--364.  Hall, Hao Ying, Langari and Yen, eds., IEEE,
  New York.

\item Jain,~A.K., and Dubes,~R.C. 1988. {\em Algorithms for
    Clustering Data}. Prentice--Hall, Englewood Cliffs.
  
\item Kamata,~S., Eason,~R.O., Kawaguchi,~E. 1991.  ``
  Classification of Landsat image data and its evaluation using a
  neural network approach'', {\em Transactions of the Society of
    Instrument and Control Engineers} {\bf 27}(11), 1302--1306.

\item Kamgar-Parsi,~B., Kanal,~L.N. 1985. ``An improved branch 
  and bound algorithm for computing k--nearest neighbors,
  {\em Pattern recognition letters} {\bf 3}, 7--12.
  
\item Kamata,~S., Kawaguchi,~E., Niimi,~M. 1995.  ``An interactive
  analysis method for multidimensional images using a Hilbert curve'',
  {\em Systems and Computers in Japan}, 83--92.

\item Kandel,~D., and Domany,~E. 1991. ``General cluster Monte Carlo
  Dynamics'', {\em Physical Review} {\bf B 43}, 8539--8548.
  
\item Karayiannis,~N.B. 1994. ``Maximum entropy clustering algorithms
  and their application in image compression'', proceedings of the
  {\em 1994 IEEE International Conference on Systems, Man, and
    Cybernetics. Humans, Information and Technology} {\bf 1}, 337--342.
  IEEE, New York.

\item Kelly,~P.M., and White,~J.M., 1993.  ``preprocessing
  remotely--sensed data for efficient analysis and classification''.
  {\em SPIE Applications of artificial intelligence 1993: knowledge
    based systems in aerospace and industry.} 24--30.
  
\item Kirkpatrick,~S., Gelatt Jr.,~C.D., and Vecchi,~M.P. 1983.
  ``Optimization by simulated annealing'', {\em Science} {\bf 220},
  671--680.

\item Kosaka,~T., and Sagayama,~S. 1994.  ``Tree--structured speaker
  clustering for fast speaker adaptation'', proceedings of the {\em
    1994 IEEE International Conference on Acoustics, Speech and Signal
    Processing} {\bf 1}, 245--248.  IEEE, New York.
  
\item   Larch,~D. 1994. `` Genetic algorithms for terrain categorization of Landsat'',  Proceedings of the {\em SPIE -- The International Society for
  Optical Engineering} {\bf 2231}, 2--6.

\item MacQueen,~J. 1967. ``Some methods for classification and
  analysis of multivariate observations'', proceedings of the {\em
    Fifth Berkeley Symp.\ Math.\ Stat.\ Prob.} {\bf I}, 281--297.
  
\item M\'{e}zard,~M., and Parisi,~G. 1986. ``A replica analysis of the 
  traveling salesman problem'', {\em Journal de Physique} {\bf 47},
  1285--1286.

\item Miller,~D., and Rose,~K. 1996. ``Hierarchical, unsupervised
  learning with growing via phase transitions'', {\em Neural
    Computation}, {\bf 8}, 425--450.
  
\item Moody,~J., and Darken,~C.J. 1989.  ``Fast learning in neural
  networks of locally--tuned processing units'', {\em Neural
    Computation} {\bf 1}, 281--294.

\item Murphy,~P.M., and Aha. ~D.W. (1994). {\em UCI repository of
    machine learning databases},
  http://www.ics.edu/\~~\hspace{-0.1cm}mlearn/MLRepository.html,
  Irvive, CA: University of California, Department of Information and
  Science.
  
\item Niedermayer,~F. 1990. ``Improving the improved estimators in O(N)
  spin models'', {\em Physical Letters B} {\bf 237}, 473--475.
  
\item Nowlan,~J.S., and Hinton,~G.E. 1991. ``Evaluation of adaptive
  mixtures of competing experts'', {\em Advances in Neural Information
    Processing Systems} {\bf 3}, 774--780.  Lippmann, Moody and
  Touretzky, eds., Morgan--Kaufmann, San Mateo.

\item Phillips,~W.E., Velthuizen,~R.P., Phuphanich,~S., Hall,~L.O.,
 Clarke,~L.P., and Silbiger,~M.L. 1995.  ``Application of
  fuzzy c--means segmentation technique for tissue differentiation in
  MR images of a hemorrhagic glioblastoma multiforme'', {\em Magnetic
    Resonance Imaging} {\bf 13}, 277--290.

\item  Rose,~K., Gurewitz,~E., and Fox,~G.C. 1990.
  ``Statistical mechanics and phase transitions in clustering'', {\em
    Physical Review Letters} {\bf 65}, 945--948.
  
\item Srinivasan A. 1994.  UCI Repository of machine learning
  databases (University of California, Irvine) maintained by P.~Murphy
  and D.~Aha (1994).

\item Suzuki,~M., Shibata,~M., and Suto,~Y. 1995.  `` Application to
  the anatomy of the volume rendering--reconstruction of sequential
  tissue section for rat's kidney'', {\em Medical Imaging
    Technology} {\bf 13}, 195--201.
  
\item Swendsen,~R.H., Wang,~S., and Ferrenberg,~A.M. 1992.  ``New
  Monte Carlo methods for improved efficiency of computer simulations
  in statistical mechanics'' in {\em The Monte Carlo method in
    condensed matter physics}, Topics in applied physics {\bf 71}, 75--91.
  Edited by K.~Binder, Springer--Verlag, Berlin.

\item Swendsen,~R.H. 1993. ``Modern methods of analyzing Monte Carlo
  computer simulations'', Physica A {\bf 194}, 53--62.

\item Wang,~S., and Swendsen,~R.H. 1990. ``Cluster Monte Carlo
  algorithms'', Physica A {\bf 167}, 565--579.
 
\item Wiseman,~S., Blatt,~M., and Domany,~E. 1996. Preprint.

\item Wolff,~U. 1989. ``Comparison between cluster Monte Carlo algorithms
  in the Ising spin model'', Physics Letters {\bf B 228}, 379--382. 
  
\item Yuille,~A.L., and Kosowsky,~J.J. 1994.  Neural Computation {\bf
    6}, 341--356.
  
\item Wong,~Y-F. 1993. ``Clustering data by melting'', {\em Neural
    Computation} {\bf 5} 89--104.
  
\item   Wu,~F.Y., 1982. ``The Potts model'',  
  {\em Reviews of modern physics}, {\bf 54}(1)235--268.
 
\end{description}

\end{document}